 \documentclass[10pt,preprint]{aastex}  
 \usepackage{natbib}
\def\gapprox{\lower.4ex\hbox{$\;\buildrel >\over{\scriptstyle\sim}\;$}}
\def\lapprox{\lower.4ex\hbox{$\;\buildrel <\over{\scriptstyle\sim}\;$}}

\shortauthors{ASCHWANDEN and GOPALSWAMY}
\shorttitle{Aerodynamic Drag in CMEs}

\begin{document}

\title{Global Energetics of Solar Flares: 
		VII. Aerodynamic Drag in Coronal Mass Ejections}

\author{        Markus J. Aschwanden$^1$}
\affil{         $^1)$ Lockheed Martin,
                Solar and Astrophysics Laboratory,
                Org. A021S, Bldg.~252, 3251 Hanover St.,
                Palo Alto, CA 94304, USA;
                e-mail: aschwanden@lmsal.com }

\and
\author{	Nat Gopalswamy$^2$}
\affil{		Heliophysics, NASA Goddard Space Flight Center,
		Greenbelt, MD 20771, USA;
		e-mail: nat.gopalswamy@nasa.gov }

\begin{abstract}
The free energy that is dissipated in a magnetic reconnection
process of a solar flare, generally accompanied by a coronal mass 
ejection (CME), has been considered as the ultimate energy source
of the global energy budget of solar flares in previous statistical 
studies. Here we explore the effects of the aerodynamic drag force
on CMEs, which supplies additional energy from the slow solar wind 
to a CME event, besides the magnetic energy supply. For this
purpose we fit the analytical aerodynamic drag model of Cargill (2004)
and Vrsnak et al.~(2013) to the height-time profiles $r(t)$ of
LASCO/SOHO data in 14,316 CME events observed during the
first 8 years (2010-2017) of the SDO era (ensuring EUV coverage
with AIA). Our main findings are:
(i) a mean solar wind speed of $w=472 \pm 414$ km s$^{-1}$,
(ii) a maximum drag-accelerated CME energy of $E_{drag} \lapprox
2 \times 10^{32}$ erg, (iii) a maximum flare-accelerated CME energy
of $E_{flare} \lapprox 1.5 \times 10^{33}$ erg; (iv) the ratio of
the summed kinetic energies of all flare-accelerated CMEs to the 
drag-accelerated CMEs amounts to a factor of 4; (v) the inclusion 
of the drag force slightly lowers the overall energy budget of CME
kinetic energies in flares from $\approx 7\%$ to $\approx 4\%$; 
and (vi) the arrival times of CMEs at Earth can be predicted 
with an accuracy of $\approx 23\%$. 
\end{abstract}


\section{	INTRODUCTION					}

The motivation for this study is the determination of the 
energy budget of {\sl coronal mass
ejections (CMEs)} in the overall global energetics and energy partitioning 
of solar flare/CME events. Previous statistical work on flare energies 
has been pioneered by Emslie et al.~(2004, 2005, 2012), and has been 
focused to the dissipation of magnetic energies (Aschwanden et al. 2014),
thermal energies (Aschwanden et al.~2015), non-thermal energies
(Aschwanden et al.~2016), CMEs (Aschwanden 2016, 2017), and the global 
energy closure between these various forms of energies 
(Aschwanden et al.~2017). It goes without saying that we cannot claim to 
understand the physics of flares and CMEs if we cannot pin down the relative
amounts of energies in such a way that we obtain closure in the total
energy budget. In the big picture we assumed that the magnetic free energy
(which is defined as the difference between the non-potential and potential
magnetic energy) provides the ultimate source and upper limit of energy that 
can be dissipated during a flare/CME event, most likely to be driven by a
magnetic reconnection process. Consequently, the potential gravitational
force and the kinetic energy of a CME have to be supplied entirely by
the magnetic free energy and the associated Lorentz forces, in addition to
the energy needed for the acceleration of particles and direct heating 
of the flare plasma. In the meantime
it became clear that additional energy (besides the dissipated magnetic
energy) can supply part of the CMEs kinematics, in form of the aerodynamic 
drag force that is exerted onto CMEs from the ambient slow solar wind
(Vrsnak and Gopalswamy 2002; Cargill 2004; Vrsnak et al.~2008, 2010, 2013). 
For a brief review see Aschwanden (2019, Section 15.5).
The main focus of this study is
therefore the question to what extent the presence of the aerodynamic
drag force affects the energy partition ratios of flare/CME events,
compared with previous studies where this effect was not taken into account.

The role of the aerodynamic drag force on coronal mass ejections (CMEs)
and interplanetary coronal
mass ejections (ICME) has been brought to recent attention (Cargill 2004;
Chen 1997; Gopalswamy et al.~2001a). Cargill (2004)
demonstrated that tenuous ICMEs rapidly equalize in velocity due to the 
very effective drag force, while ICMEs that are denser than the ambient 
solar wind are less affected by the aerodynamic drag, although the drag 
coefficient is approximately independent of the propagation distance. 
An anti-correlation between the CME acceleration and velocity was established
from LASCO/SOHO data (Gopalswamy et al.~2000, 2001a),
which confirms that massive CMEs are less affected 
by the aerodynamic drag (Vrsnak et al.~2008).
Massive CMEs have been found to be accelerated for masses of 
$m_{cme} > 3\times 10^{14}$ g, while less massive CMEs are generally decelerated
(Michalek 2012).
The shortest transit times and hence the fastest velocities have been
identified in narrow and massive ICMEs (i.e., high-density eruptions)
propagating in high-speed solar wind streams 
(Gopalswamy et al.~2000, 2001a; Vrsnak et al.~2010).
Extremely short transit times of 
14 hours (Gopalswamy et al.~2005a) and
21 hours have been observed (Temmer and Nitta 2005),
with maximum speeds of $v \approx 2600$ km s$^{-1}$,
but agreement with the aerodynamic drag model
requires a decrease of the solar wind density near 1 AU
(Temmer and Nitta 2015), but see Gopalswamy et al.~(2016) for
an alternative interpretation.
Fast CMEs were found to show a linear dependence for the velocity
difference between CMEs and solar wind, while slow CMEs show a
quadratic dependence (Maloney and Gallagher 2010). 
A quadratic dependence is expected in a collisionless environment,
where drag is caused primarily by emission of magnetohydrodynamic (MHD)
waves (Vrsnak et al.~2013). A distinction between the aerodynamic drag
force and the hydrodynamic Stokes drag force has been suggested
(Iju et al.~2014), but was found to be equivalent in other
cases (Gopalswamy et al.~2001b).
Analytical models for the drag coefficient include the viscosity
in the turbulent solar wind (Subramanian et al.~2012).
The aerodynamic drag model has been used increasingly as the preferred 
physical model to quantify the propagation of ICMEs and to forecast their 
arrival times at Earth, and this way it became a key player in space weather 
predictions (Michalek et al.~2004; Vrsnak et al.~2010; Song 2010; 
Shen et al.~2012; Kilpua et al.~2012; Lugaz and Kintner 2013;
Hess and Zhang 2014 ; Tucker-Hood et al.~2014; Mittal and Narain 2015; 
Zic et al.~2015; Sachdeva et al.~2015; Dumbovic et al.~2018;
Verbeke et al.~2019).
Arrival times at Earth inferred from the ``drag-based model'' have
been compared with the numerical ``WSA-ENLIL+Cone model''
(Wang-Sheeley-Arge), which
enables early space-weather forecast 2-4 days before the arrival of the
disturbance at Earth (Vrsnak et al.~2014; Dumbovic et al.~2018). 
The Stokes form was the basis for the empirical shock arrival
model, whose prediction is comparable to that of the 
ENLIL+cone model (Gopalswamy et al.~2005b, 2013).
New models, such as the {\sl Forecasting a CMEs Altered Trajectory
(ForeCAT)} deal also with CME reflections based on magnetic forces
and non-radial drag coefficients (Kay et al.~2015).
Geometric models, such as the {\sl Graduated Cylindrical Shell (GCS)}
model are fitted to LASCO and STEREO data, finding that the Lorentz
forces generally peak at $(1.65-2.45) R_{\odot}$, and become
negligible compared with the aerodynamic drag already at distances of 
$(3.5-4.0) R_{\odot}$, but only at $(12-50) R_{\odot}$ for slow
CME events (Sachdeva et al.~2017).

In this paper we are fitting the aerodynamic drag model to all CMEs
observed with LASCO/SOHO during the SDO era (2010-1017), which
yields the physical parameters that are necessary to determine
the kinetic energies, the energy ratios of flare-associated and
drag-accelerated CMEs, as well as their arrival times near Earth.
We present the analytical description of the constant-acceleration
and aerodynamic drag model in Section 2, the data analysis of
forward-fitting the analytical models to LASCO data and the
related results in Section 3, a discussion of some relevant issues
in Section 4 and conclusions in Section 5. 

\section{	THEORY AND METHODS 				}

\subsection{	The Constant-Acceleration Model			}

The simplest model of the kinematics of a coronal mass ejection 
(CME) has a minimum number of three free parameters, which includes 
a constant (time-averaged) acceleration $a_0$, an initial height 
$r(t=t_0)=r_0$, and a starting time at a reference time $t=t_0$. 
A slightly more general model (with 4 free parameters) allows also 
for a non-zero velocity $v_0=v(t=t_0)$ at the starting time $t=t_0$,
which constitutes four free model parameters $[a_0, v_0, r_0, t_0]$,
defining the time dependence of the acceleration $a(t)$,
\begin{equation}  
	a(t)=a_0 \ ,
\end{equation}
the velocity $v(t)$ of the CME leading edge,
\begin{equation}  
	v(t)=\int_{t_0}^t a(t) dt = v_0 + a_0 (t-t_0) \ ,
\end{equation}
and the radial distance $r(t)$ from Sun center,
\begin{equation}  
	r(t)=\int_{t_0}^t v(t) \ dt = 
	r_0 + v_0 (t-t_0) + {a_0 \over 2} (t-t_0)^2 \ .
\end{equation}
The radial distance $r(t)$, which is directly obtained from
the observations, can be fitted with a simple second-order
polynomial,
\begin{equation}
	r(t) = c_0 + c_1 t + c_2 t^2 \ ,
\end{equation}
where the free parameters as functions of the coefficients 
$c_0, c_1, c_2$ follow directly from Eqs.~(3) and (4),
\begin{equation}
	a_0 = 2 c_2 \ ,
\end{equation}
\begin{equation}
	t_0 = (v_0 - c_1)/a_0 \ ,
\end{equation}
\begin{equation}
	r_0 = c_0 + v_0 t_0 - {a_0 \over 2} t_0^2 \ .
\end{equation}
A practical example of the height-time profile $r(t)$, the
velocity profile $v(t)$, and the acceleration profile $a(t)$
of the constant-acceleration CME kinematic model is shown  
for an event in Fig.~1 (left), where the observed 
datapoints $r_i=r(t=t_i)$
are marked with crosses (top left panel), and the fitted model
is rendered with thick curves, covering the fitted time range 
$[t_1, t_2]$. For the fitting of an acceleration model
to LASCO data we have to be aware that CMEs are observed at a
heliocentric distance of $\gapprox 2.5 R_{\odot}$, by which time most
CMEs have finished acceleration (Bein et al.~2011) and we are
observing a residual acceleration only, combined with gravity
and drag.

For the calculation of the CME starting time $t_s$ we extrapolate
the model $r(t)$, which is observed in the time range $[t_1, t_2]$,
to an expanded range $[t_0,t_2]$ with double length (with lower boundary 
$t_0=t_1-(t_2-t_1)$). The actual starting time $t_s$ of the CME 
launch can now be derived from the
height-time profile $r(t)$ within the expanded time range
$[t_0, t_2]$, where two possible cases can occur. One case is 
when the extrapolated minimum height at the start of the CME 
is lower than the solar limb, in which case the solution $r(t)$ 
can simple be extrapolated to the nominal height $r_s = 1 R_{\odot}$,
as it is shown for the case depicted in Fig.~1 (left), for the
CME event on 2011 September 24, 18:36 UT.

The other case is when the minimum height 
$r_s=min[r(t)]=r(t=t_s)$ is higher than the solar limb 
$(r_{min} > R_{\odot})$, in which case the starting time $t_s$ 
coincides with the height minimum, where the velocity is zero, i.e., 
$v_s=v(t=t_s)=0$. An example of the second case is shown in Fig.~2 (left),
where the starting height is estimated to $r_s=1.963 R_{\odot}$
for the event of 2010 January 3, 05:30 UT.
Note that the starting
time, defined by the extrapolated zero velocity $v_s(t=t_s)=0$,
is dependent on the model, estimated at $t_s=2.375$ hrs for the
constant-acceleration model, and $t_s=6.287$ hrs for the
aerodynamic drag model. So there is an uncertainty of the order
of $\approx 4$ hrs for the start of this particular event.

\subsection{	The Aerodynamic Drag Model 			}

We define now, besides the constant-acceleration model, a second model
that is based on a physical mechanism.
The interaction of a coronal mass ejection (CME)
(or an interplanetary coronal mass ejection (ICME))
with the solar wind leads to an adjustment or equalization
of their velocities at heliocentric distances from a few solar 
radii out out to one astronomical unit.  When an ICME has initially 
a higher velocity 
(or take-off speed) than the solar wind, it is then slowed down to 
a lower value that is closer to the solar wind speed (Fig.~1). Vice versa,
ICMEs with slower speeds than the ambient solar wind speed become 
accelerated to about the solar wind speed (Fig.~2). 
There are also cases where the CME accelerates to high speeds,
but quickly slows down even before the solar wind formation
(Gopalswamy et al.~2012; 2017). Following the physical 
model of aerodynamic drag formulated by Cargill (2004) and the 
analytical solution of Vrsnak et al.~(2013), we can describe the 
velocity time profile $v(t)$,
\begin{equation}
	v(t) = \left( { dr(t) \over dt } \right) 
	= {(v_s - w) \over
		1 \pm \gamma (v_s - w) (t - t_s) } + w  \ ,
\end{equation}
where $v_s$ is the CME velocity at an initial start time $t_s$
(also called ``take-off'' velocity), $w$ is the (constant) solar
wind speed, $\gamma \approx 1 \times 10^{-7}$ cm$^{-1}$ is the drag
parameter (in units of inverse length),
and $r_s$ is the initial height at the starting time $t=t_s$. 
The drag parameter $\gamma$ has been defined as
\begin{equation}
	\gamma = {c_d A \rho_w \over M + M_v} \ ,
\end{equation}
where $c_d$ is the dimensionless drag coefficient (Cargill 2004),
$A$ is the ICME cross-sectional area, $\rho_s$ is the ambient
solar-wind density, and $M$ is the ICME mass. The so called
virtual mass, $M_v$, can be expressed approximately as
$M_v \approx \rho_w V/2$, where $V$ is the ICME volume.
Here we assumed a constant solar wind speed $w$ and a constant $\gamma$,
which is justified to some extent by MHD simulations (Cargill 2004),
which show the constant drag coefficient $c_d$ varies slowly
between the Sun and 1 AU, and is of order unity. When the ICME and
solar wind densities are similar, $c_d$ becomes larger, but
remains approximately constant with radial distance. For ICMEs
denser than the ambient solar wind, $\gamma$ is approximately independent
of radius, while $\gamma$ falls off linearly with distance 
for tenuous ICMEs (Cargill 2004). Regarding the variability of the
solar wind speed $w(r)$ as a function of the distance $r$, 
the largest deviation from a constant value $w(r)$ is expected in the
corona, where the solar wind transitions from subsonic to supersonic
speed at a distance of a few solar radii from Sun center, but this
coronal zone is also the place where flare-associated acceleration
of CMEs occurs and aerodynamic drag is less dominant, which 
alleviates the influence of the (non-constant) solar wind.

Vrsnak et al.~(2003) integrated the velocity dependence
$v(t)$ (Eq.~8) to obtain an analytical function for the
height-time profile $r(t)$ explicitly,
\begin{equation}
	r(t) = \pm {1 \over \gamma}
	\ln{ [ 1 \pm \gamma (v_s - w) (t - t_s) ] }
		+ w (t - t_s) + r_s \ .
\end{equation} 
Differentiating the speed $v(t)$, we obtain
an analytical expression for the acceleration time profile $a(t)$,
\begin{equation}
	a(t) = \left({ dv(t) \over dt } \right) = 
		{ \mp \gamma (v_s - w)^2 \over
		[1 \pm \gamma (v_s - w) (t - t_s)]^2} \ .
\end{equation}
We see that this model has 5 free parameters $[t_s, v_s, r_s, w, \gamma]$.
The two regimes of $\pm$ correspond to the deceleration/acceleration
regime, i.e., it is plus for $v_s > w$, and minus for $v_s < w$.
Comparing with the constant-acceleration model, we see that three
parameters are equivalent, i.e., $[t_s, r_s, v_s]=
[t_0, r_0, v_0]$, while the acceleration $a_0$ is constrained by the
drag coefficient $\gamma$ and the solar wind speed $w$.

Two examples of CME kinematic models with the aerodynamic drag
model are shown in Figs.~1 and 2 (right-hand panels).
The case shown in Fig.~1 (middle right panel) reveals deceleration from
and initial value of $v_s = 801$ km s$^{-1}$ towards the solar wind
speed of $w = 405$ km s$^{-1}$, while the other case 
shows acceleration from $v_s = 0$ km s$^{-1}$ to $w = 421$ km s$^{-1}$,
according to the aerodynamic drag model (Fig.2, middle right panel).

For the calculation of the free parameters 
we define a fitting time range $[t_s,t_2]$ that is bound by the 
starting time $t_s$ of the CME (inferred from the constant-acceleration 
model) and the last observed time $t_2$ of the LASCO/SOHO data.
The remaining four free parameters $[r_s, v_s, w, \gamma]$ are optimized 
by forward-fitting of the height-time profile $r(t)$ (Eq.~10) to
the observed heights $r_i=r(t_i), i=0,...,n_t$ of the LASCO/SOHO data,
using the {\sl Direction Set (Powell's)} methods in multidimensions
(Press et al.~1986). A robust performance of this optimization
algorithm is achieved by optimizing the parameters [$ln(w)$, $\ln(v_s/w)$, 
$\ln{(\gamma)}$, $r_s/R_{\odot}$]. The iteration of logarithmic parameters
avoids (unphysical) negative values for the velocities and the drag
parameter, [$v_s, w, \gamma$].
 
\section{	OBSERVATIONS AND DATA ANALYSIS RESULTS 		}

\subsection{	LASCO/SOHO Data}

In the following we describe the observations from LASCO/SOHO 
and characterize the statistical results of our data analysis. 
We make use of the SOHO/LASCO CME catalog that is publicly 
available at {\sl https://cdaw.gsfc.nasa.gov/CME$\_$list},
based on visually selected CME events,
created and maintained by Seiji Yashiro and Nat Gopalswamy
(Yashiro et al.~2008; Gopalswamy et al.~2009a, 2010).
A brief description of the algorithm of measuring height-time
profiles $r(t)$ is given on the same website. From the
LASCO/SOHO data archive, only C2 and C3 data have been used 
for uniformity, because LASCO/C1 has been disabled in June 1998. 
We downloaded the time sequences of height time profiles, 
$r_i = r(t=t_i), i=1,...,n_t$, 
that are available for every CME detected with LASCO/SOHO during
the first 8 years (2010-2017) of the {\sl Solar Dynamics Observatory 
(SDO)} mission. This data set comprises 14,316 events, covering
almost a full solar cycle.

\subsection{	Fitting of CME Kinematic Models	}

The forward-fitting of both the constant-acceleration model
(Section 2.1) and the aerodynamic drag model (Section 2.2)
to the LASCO height-time profiles yields dynamical parameters 
that are important for extrapolating the CME kinematics from 
the LASCO-covered distance range of $r \approx (3 - 32) R_{\odot}$ 
to the lower corona at $r \lapprox 1.5 R_{\odot}$ (for 
identification of simultaneous flare events), and extrapolating
out into the heliosphere to $r \approx 1$ AU 
(for forecasting of the CME arrival time at Earth). 

We fitted the constant-acceleration
model (Section 2.1) to the LASCO/SOHO CME height-time profiles
in the same way as the second-order polynomial fits have been 
carried out in the LASCO CME catalog, and we verified consistency
between our fits and those listed in the CDAW LASCO CME catalog. 
We found that the forward-fitting of both models is fairly robust.
Unsatisfactory fits have been found in very few cases, identified 
by a low fitting accuracy ($\sigma \lapprox 5\%$, in 6\% of the
cases for the constant-acceleration model, and in 7\% of the cases
for the aerodynamic drag model), or a low drag coefficient 
($\gamma \le 10^{-8}$ cm$^{-1}$, in 10\% of the cases).

The fitting quality of the two (analytical) theoretical models 
used here is defined as follows. We calculate the average
ratios of the fitted (modeled) distances $r_i^{model}$ and 
compare them with the observed distances, $r_i^{obs}=r(t_i), i=1,...,n_t$, 
\begin{equation}
	q = {1 \over n_{t}} \sum_{i=1}^{n_t} 
	    \left( {r_i^{model} \over r_i^{obs}} \right)
	    \approx 1.0 \pm \sigma \ .
\end{equation}
This measure of the accuracy has been found to be very suitable,
yielding a standard deviation of $\sigma_{CA}=2.7\%\pm2.7\%$ 
for the constant-acceleration model,
and a very similar value of $\sigma_{AD}=2.9\%\pm2.5\%$ for the
aerodynamic drag model. The accuracies were calculated
for all 14,316 events, based on an average of $n_t \approx 23$ 
distance measurements per event. The fact that both models fit
the data with equal accuracy suggests that either model is
suitable. The example shown in Fig.~1 reveals an equal accuracy 
of $\sigma_{AC}=\sigma_{AD}=4.4\%$ for both models. 
The example shown in Fig.~2 yields a better performance for the
constant-acceleration model ($\sigma_{CA}=3.3\%$), versus
$\sigma_{AD}=5.4\%$ for the aerodynamic drag model. 
However, the aerodynamic drag model 
represents a physical model and yields the five
parameters $[t_s, r_s, v_s, w, \gamma]$, while the 
constant-acceleration model requires four parameters 
$[t_0, r_0, v_0, a_0]$. A fundamental difference between the
two models is that the acceleration $a_0$ is not time-dependent in the 
constant-acceleration model, while it is variable in the
aerodynamic drag model, and the CME speed asymptotically approaches 
the solar wind speed constraining the solar wind speed $w$ and 
the aerodynamic drag coefficient $\gamma$.

\subsection{	Eruptive and Failed CMEs 	} 

A distinction is generally made by the dynamical characteristics
of CME events, which defines the type of {\sl eruptive flares 
or CMEs} when the final CME speed exceeds the gravitational 
escape velocity ($v_2 \ge v_{esc}$), and alternatively the 
type {\sl failed eruptions}, when the escape velocity is not reached 
($v_s < v_{esc}$). Failed eruptions have mass motions, but do not escape
(Gopalswamy et al.~2009b).
The escape speed depends only on the radial distance from the 
Sun center,
\begin{equation}
	v_{esc}(r) = \sqrt{ {2 G M_{\odot} \over r} }
		\approx 618\ 
		\left( {r  \over R_{\odot}} \right)^{-1/2} 
		{\rm km\ s}^{-1} \ ,
\end{equation}
where $G$ is the gravitational constant, $M_{\odot}$ the
solar mass, and $R_{\odot}$ the solar radius. Examples
of the escape speed dependence on the radial distance
are shown in Figs.~1 and 2 (middle panels), where the
escape speed is indicated with dotted curves. In the first
event, the CME speed exceeds the escape velocity 
all the times (Fig.~1 middle right),
while the second case reaches escape speed at
$t = 12.1$ hrs (Fig.~2, middle right), which is 
reached at a distance of $r = 7.2 R_{\odot}$ 
(Fig.~2 top right). So, both events are eruptive CMEs. 

We determined the time $t_{esc}$ and distance $r_{esc}$ 
where the CME gained sufficient speed to overcome the
combined Lorentz force, gravitational force, and 
drag force, as a function of the time (Fig.~3a)
and as a function of the distance (Fig.~3b), for all 
analyzed 14,316 CME events. The start time $t_s$ has been
extrapolated from the aerodynamic drag model to a starting
height of $r_s \approx 1 R_{\odot}$. We see that $\approx 10\%$
of the CMEs reach escape velocity below the solar limb (Fig.~3b), 
that $\approx 30\%$ of the CMEs reach escape velocity at 
$r_1 \lapprox 3 R_{\odot}$,
at the location of the LASCO first detection, and that 100\%
reach escape speed at a distance of $r_2 \lapprox 10 R_{\odot}$,
at the location of the LASCO last detection. Conversely, all CMEs
reach escape velocity at $\lapprox 25$ hrs after launch
(Fig.~3a). This confirms the selection criterion of the
CDAW CME list, where eruptive CME events have been
measured only, by definition. No confined flare is contained in 
the CDAW CME list, but we will encounter such events when we
compare the association of soft X-ray flare events with CME 
detections, using GOES flare data (see Section 3.5).  

\subsection{	Statistical Results of LASCO Fitting 		}

We summarize the statistical results of our fitting of the two
CME kinematic models (Eqs.~3 and 10) in Table 1 and in the
Figures~4 to 7. 
In Fig.~4 we show the near-final speed $v_2$ (that is measured from 
the last detection in LASCO data), versus the ambient slow solar wind 
speed $w$ 
for all CME events. Although the last detection with LASCO yields 
a wide range of final speeds $v_2 \approx 100-1000$ km s$^{-1}$ 
(Fig.~4), the slow slow solar wind is mostly
concentrated in the velocity range of $w \approx 200-500$ 
km s$^{-1}$. This implies that the aerodynamic drag model 
accelerates CMEs with $v_s < w$ and decelerates CMEs with
$v_s > w$ towards the near-final speed of $v_2 \approx 200-500$
km s$^{-1}$, as indicated with the concentration of data
along the vertical ridge of $w \approx 400$ km s$^{-1}$ (Fig.~4).
This agrees also with the conclusion obtained from the 
empirical acceleration formula derived by Gopalswamy et al.~(2001b),
i.e., $a=-0.0054 (v_{cme} - 406)$.

This confirms that the solar wind speed $w$ is reliably retrieved 
from forward-fitting of the kinematic model (Eq.~8) to the LASCO 
data, regardless what the value of the CME speed $v_2$ is.
The most frequent starting height is in the lower corona, 
at a median distance of $r_s \lapprox 1.2 R_{\odot}$ 
from Sun center ($h_S \lapprox 140,000$ km)
(Fig.~5a, Table 1). 
The first detection with LASCO occurs at a mean distance of
$r_1 = (3.0 \pm 0.8) R_{\odot}$ (Fig.~5b), while the last detection
with LASCO is around $r_2 = (10.3 \pm 6.4) R_{\odot}$ 
(Fig.~5c, Table 1). 

Statistics of the velocities are particularly interesting here
because the propagation of most CMEs depends on their relative
speed to the slow solar wind speed. The distribution of
starting speeds $v_s$ has two peaks (Fig.~6a), one near
$v_s \approx 0$, and a second peak at $v_s \approx 200$ km s$^{-1}$.
This bimodality depends on the time resolution, which is typically
0.2 hrs or 12 minutes (Table 1) for LASCO data. 
If the initial acceleration of CMEs in the lower solar corona
peaks before 12 minutes, we do not resolve the initial speed
increase from $v_s=0$ to $v_1 \approx 200$ km s$^{-1}$, while
a peak acceleration later than 12 minutes after the starting
time $t=t_s$ will reveal an initial value of $v_s \approx 0$
(e.g., Fig.~2). This explains the large range of obtained 
starting velocities $v_s=482 \pm 1294$ km s$^{-1}$,
which has a higher mean than that at the first detection with LASCO,
$v_1 = 320 \pm 283$ km s$^{-1}$ (Fig.~6b), or at the last detection with LASCO,
$v_2 = 368 \pm 198$ km s$^{-1}$ (Fig.~6c), due to the aerodynamic drag that
streamlines the CME velocities.

The most interesting statistical result is the distribution of
slow solar wind speeds, which have an average of $w=472\pm414$ km s$^{-1}$
or a median of $w=405$ km s$^{-1}$ (Fig.~6d),
obtained at heliocentric distances in the range of
$r_2 \approx 3-30 R_{\odot}$, according to the last LASCO detection
(shown in Fig.~5c).
Note that these forward-fitting results of the aerodynamic drag model, 
based on 14,306 LASCO CME events, represent one of the largest
statistical measurements of the slow solar wind speeds. 

We present the statistical time scales related to the LASCO
detection delay $(t_1-t_s)$ and propagation duration in the
LASCO field-of-view in Fig.~7 and Table 1. The average 
detection delay is $(t_1-t_s)=1.0 \pm 1.3$ hrs (Fig.~7a).
The mean duration of CME propagation in the LASCO observed zone
is $(t_2-t_1)=4.3 \pm 3.7$ hrs (Fig.~7b).

\subsection{	CME Start Times and GOES Flare Times	}

While the previously described results make exclusively
use of LASCO/SOHO data, we compare now these measurements with
other data sets based on HMI/SDO, AIA/SDO, and GOES data.
In particular we focus on a subset of 576 M and X-class GOES
flare events that have been observed during the first 7 years
of the SDO mission (2010-2016), for which 
measurements of temporal, spatial, and energetic parameters 
were published previously (Aschwanden 2016, 2017).

In order to identify LASCO CME events that are associated with
each of the 576 M and X-class GOES flares we use the GOES flare
start reference times $t_s^{GOES}$ issued by NOAA, and find the 
CME events (of the entire LASCO catalog of 14,316 events during 
2010-2017) that have their first LASCO detection time $t_1^{LASCO}$
closest to the starting time $t_s^{GOES}$ of the GOES flares. The 
relative time difference can be significantly improved by 
extrapolating the LASCO height-time plot $r(t)$ to the 
LASCO starting time $t_s^{LASCO}$ at the initial height of
$r_s=r(t=t_s)$, which yields a time difference between the
GOES and LASCO starting times,
\begin{equation}
	\Delta t = t_s^{LASCO} - t_s^{GOES} \ .
\end{equation}
A histogram of the time differences between the GOES start times
$t_s^{GOES}$ and the extrapolated LASCO starting times $t_s^{LASCO}$
is shown in Fig.~8. Out of the 576 events we find a total of 480 
events (83\%) with relative time delays within a time window of $\pm 4$ 
hrs. We calculate a Gaussian fit to the core distribution within
a time window of $\pm0.7$ hrs, which encompasses 231 events (40\%)
that may be considered as a lower limit of events with good time 
coincidence. 
The association rate of a CME with a flare increases from 
20\% for C-flares to 100\% in large X-class events (Yashiro et al.~2005).
Therefore, the flare-associated fraction of LASCO
CME events may vary between the limits of 20\% and 100\%.  
On the other hand, the complementary fraction of GOES flare events that
have no CME detected with LASCO may vary in the range of 17\% to 60\%.
These CME-less events that are not associated with a $>$M.1 GOES class 
flare may consist of confined flares or weak non-detected CME events.

The distribution of starting delays is
$\tau = t_s^{LASCO}-t_s^{GOES} = 0.07 \pm 0.28$ hrs (Fig.~8),
evaluated with a Gaussian fit at the peak of the distribution.
This is consistent with previous measurements of 275 flare/CME events, 
where also no significant delay was found, i.e., 
$\tau = 0.02 \pm 0.77$ hrs (Fig.~17c in Aschwanden 2016).
In this relative timing analysis we neglected the difference
of the heliographic position of CME source locations, 
since the propagation time difference from disk center to the limb,
i.e., $\Delta t_{prop}=R_{\odot}/v_{cme} \approx 0.2$ hrs (for
$v_{cme} \approx 1000$ km s$^{-1}$) is smaller than the bin
width of the histogram shown in Fig.~8. 

Four examples of flare/CME events are shown in form of GOES flux
time profiles and CME height-time plots (Fig.~9), which illustrate different
uncertainties in the time coincidence. The first example (Fig.~9a)
exhibits a simple single-peak GOES time profile, where the 
GOES and LASCO starting times coincide within 0.174 hrs (or 10 minutes).
The second example (Fig.~9b) has an extremely impulsive peak, but 
the extrapolated LASCO starting time has due to the low initial
speed ($v_1=71$ km s$^{-1}$ at the first detection with LASCO)
a large uncertainty, so that the coincidence is within
0.542 hrs or 33 minutes). The third case (Fig.~9c) exhibits
a substantial uncertainty in the GOES starting time, so that
the coincidence is within 0.40 hrs (or 24 minutes).
The fourth example (Fig.~9d) shows an X-class flare with a
well-defined starting time, but the initial CME speed is so low
that it implies a large negative delay of -1.56 hrs (or 94 minutes).
In summary, the accuracy of the relative
time coincidence between soft X-ray emission (detected with GOES)
and the starting height of CMEs (detected with LASCO) depends
on the definition of the flaring time (starting, peak, or end time)
and the uncertainties of the CME speed extrapolation, particularly
in the case of slow CMEs. 

Another time marker of CME starting times is the EUV dimming,
which has been measured with large statistics using AIA data.
A significant delay has been observed between the 
AIA dimming and the GOES starting time, with a mean of
$\tau = (t_s^{AIA} - t_s^{GOES}) = 0.35 \pm 0.40$ hrs (or
$21 \pm 24$ minutes) (Fig.~17d in Aschwanden 2016). 

\subsection{	Energetics of CMEs	}

Our main interest of this study is how much the 
aerodynamic drag force affects the global energetics
of flare/CME events. In order to evaluate this effect
quantitatively, we have to discriminate between the
flare-associated acceleration in the lower corona 
and the solar wind-associated acceleration in the 
heliosphere. Given a cadence of $\Delta t=12$ minutes for LASCO data,
and assuming a minimal CME velocity of $v_{min} \approx 100$ km
$s^{-1}$, the altitude range of flare-associated acceleration
is estimated to be $h_{min} = v_{min} \Delta t \gapprox 72,000$ km
or $0.1 R_{\odot}$, which corresponds to a radial distance
of $r \le 1.1 R_{\odot}$. Since the velocity corresponds to
the product of the acceleration $a$ and the acceleration time
interval $\Delta t$, i.e., $v = a \Delta t$, the absolute value of the
unresolved acceleration $a$ cannot be determined from LASCO
observations alone, but only the product. Using EUV dimming
data in addition, however, the acceleration can be resolved,
as shown from AIA/SDO data (Aschwanden 2017), where a median
acceleration rate of $a=0.8$ km s$^{-1}$, median acceleration
times of 500 s (or 7 minutes), and an acceleration height of
$h_{acc}=0.75 R_{\odot}$ have been determined (Table 1 in
Aschwanden 2017). These measurements justify the assumption
that the flare-associated acceleration occurs at low coronal 
heights of $r \lapprox 1.5 R_{\odot}$
(Gopalswamy et al.~2009; Bein et al.~2011), 
even for ground-level enhancement (GLE)
events (Gopalswamy et al.~2013). For clarification,
we emphasize that the term acceleration refers to the combination
of the Lorentz force, the gravitational force, and the drag force.

A clear indication of dominant flare-associated acceleration
is given when the CME velocity profile shows the maximum
velocity at the starting time $t_s$, i.e., $v_s=v(t=t_s)$, 
while the velocity decreases during the outward propagation,
which can be observed from the velocity difference between
the first ($v_1$) and last ($v_2$) LASCO detection, 
i.e., when $v_2 < v_1$, as used in Gopalswamy et al.~(2017),
\begin{equation} 
	E_{flare} = {1 \over 2}\ m_{cme} \left\{
		\begin{array}{cc}
		v_s^2 & \mbox{for $v_1 > v_2$} \\
		0     & \mbox{for $v_1 < v_2$} 
		\end{array}
	\right.
\end{equation}
On the other hand, the aerodynamic drag acceleration 
becomes progressively more important after the first 
detection of LASCO (at velocity $v_1$), while the last 
detection with LASCO (at velocity $v_2$) approaches the 
final CME speed, often close to the slow solar wind $w$,
\begin{equation} 
	E_{drag} = {1 \over 2}\ m_{cme} \left\{
		\begin{array}{cc}
		0     & \mbox{for $v_1 > v_2$} \\
		w^2   & \mbox{for $v_1 < v_2$} 
		\end{array}
	\right.
\end{equation}

Using these criteria we find $N_{flare}=313$ that show 
flare-associated acceleration (Fig.~10a), and $N_{drag}=263$
that exhibit aerodynamic drag acceleration (Fig.~10b),
out of the total number of 576 cases (Fig.~10c). We show
the (logarithmic) size distributions of these three data sets,
which reveal that flare-associated acceleration processes
produce the largest CME energies (Fig.~10a), while aerodynamic drag
acceleration appears to have an upper limit of $E_{drag}
\lapprox 2 \times 10^{32}$ erg (Fig.~10b, vertical dashed line).
When we integrate the CME kinetic energies over all events of
each subgroup, we find the flare-associated acceleration
processes make up for a fraction of $E_{flare}/E_{all}=80.5\%$,
while aerodynamic drag acceleration accounts for the remainder
$E_{drag}/E_{all}=19.5\%$, which obviously is dominated by the
largest events (or the most energetic CMEs). This result is
consistent with the expectation that the fastest and most
energetic CMEs are less influenced by aerodynamic drag,
because they have higher masses and higher velocities,
(although the drag fraction is larger for faster CMEs 
before the solar wind takes over).

For comparison, we show also the distribution of CME energies
in Fig.~10c (histogram with thin line style) from a previous study
(Aschwanden 2017), which has the same number of 576 events,
but contains about 1.5 times the total energy, which appears 
to be produced by a factor of 1.25 higher velocities for the
largest events at energies $E_{max} \approx 2 \times 10^{32}$ erg.
 
In summary, the energy ratio of the flare-accelerated CMEs to
the drag-accelerated CMEs is a factor of 4 for the CME kinetic energies. 
Since the CME kinetic energy accounts for 7\% of the total
flare energy budget (Aschwanden et al.~2017), 
the inclusion of the aerodynamic drag
effect lowers the CME contribution from 7\% to $7\% \times 0.8 = 5.6\%$. 
In addition, the absolute value of the CME energies is a factor 
of 1/1.5=0.67 lower in this study, which lowers the CME contribution to 
$7\% \times 0.8 \times 0.67 \approx 3.8\%$, causing an overall
change of 7\%-3.8\%=3.2\% in the global flare/CME energy budget. 

The kinetic energies of CMEs shown in Fig.~(10) have been derived
from the AIA dataset of $\approx 576$ M and X-class flares, and
thus are all associated with flares. If we ask whether flare-less
CME events have a different distribution of kinetic energies, because
they are all accelerated by the aerodynamic drag force, we would need
a data set of LASCO-detected CMEs that have no associated flares,
but heliographic flare locations are unfortunately not provided in
the LASCO CME catalog, and thus we are not able to derive kinetic
energies of events that are not associated with flares.
However, since the association rate is near 100\% for
X-class flares, we do not expect that the size distributions
shown in Fig.10 change at the upper end. For C-class flares,
however, where the flare-association rate is 20\%, we would
expect a lot of smaller CME events without flares, which
would steepen the size distributions of kinetic energies
at the lower end (Fig.~10).

\subsection{	Extrapolated CME-Earth Arrival Times	}

We may ask whether the LASCO/SOHO data (providing height-time 
series of the leading edge of propagating CMEs in a distance
range of $\approx (3-20) R_{\odot}$) are sufficient
to predict the arrival times of {\sl Interplanetary Coronal
Mass Ejections (ICME)} near Earth (e.g., Gopalswamy et al.~2013). 
Using data acquired with
the instruments onboard WIND, ACE, SOHO/CELIAS/MTOF/PM, 
we obtain timing information for the arrival times at Earth
from the ICME catalog 
http://www.srl.caltech.edu/ACE/ASC/DATA/level3/icmetable2.html
(produced by I. Richardson and H. Cane), which contains
ICME observations during 1996-2018. During the SDO+LASCO era
(2010-2017), which is of primary interest here, information
on the LASCO (or GOES) starting time are available for 78 ICME
events, whereof 19 events are associated with GOES $>M1.0$ class
flares. Eliminating events that have insufficient data points 
($n_t < 5$) or have extremely low drag coefficient values ($\gamma
\lapprox 10^{-8}$ cm$^{-1}$), we are left with
11 events, which are listed in Table 2. In Table 2 we list
the GOES flare starting times (which are good proxies for the
CME starting times $t_s$) and the ICME arrival times at
Earth, based on the time of the associated geomagnetic storm
sudden commencement, which is typically related to the arrival
of a shock at Earth (see footnotes in ICME catalog by
Richardson and Cane). The resulting observed ICME propagation
time delay $\tau_{obs}$ ranges from 35 to 87 hours (Fig.~11). 
For the predicted delay we assume radial propagation of CMEs,
which corresponds to an interplanetary path length $L_{path}$
of one astronomical unit $d_{AU}$, 
\begin{equation}
		L_{path}=d_{AU} \ ,
\end{equation}
and the mean CME speed is approximated by the last LASCO
detection $v_2$, which yields a predicted propagation delay 
$\tau_{pred}$ of 
\begin{equation}
	 	\tau_{pred} = {L_{path} \over v_2}
		= {d_{AU}\ q_{corr} \over v_2} \ ,	
\end{equation}
where the correction factor $q_{corr}$ includes various
effects that have to be determined
empirically, such as velocity corrections due to projection 
effects (since all CME velocities are measured in the 
plane-of-sky and may underestimate the true 3-D velocity,
by factors up to 2), the temporal variability of the CME 
speed, velocity changes due to CME-CME interactions,
and the temporal evolution of the solar wind speed.
We find that an empirical value of $q_{corr}=0.81$
provides the optimum correction factor.
The resulting ratio of the theoretically predicted to the observationally
measured ICME propagation delays has then a mean and standard deviation of
$\tau_{pred}/\tau_{obs}=1.00\pm0.23$ (Fig.~11a), which implies that we 
can predict the arrival times at Earth with an accuracy of $\approx 23\%$. 

Comparing our result with the empirical formula of transit
times $\tau_{pred}$ as a function of the CME speed $v$ fitted in
Gopalswamy et al.~2005a (Fig.~8 therein) for 4 ICME events, 
i.e., $\tau_{pred} = a b^v + c$ (with the best-fitting coefficients 
a=151.02, b=0.998625, c=11.5981), we find an almost identical result, 
with a mean and standard deviation of
$\tau_{pred}/\tau_{obs}=1.01\pm0.23$ (Fig.~11b).

\section{	DISCUSSION 					}

\subsection{	Eruptive and Failed CMEs			}

This classification into eruptive and failed CMEs is not trivial, 
because both the CME speed and the escape speed are spatially and 
temporally varying. In previous studies, the kinetic energy $E_{kin}(r)$ 
and the CME gravitational energy $E_{grav}(r)$ were calculated as a
function of the distance $r$ from the Sun (Vourlidas et al.~2000;
Aschwanden 2016). In the study of Vourlidas et al.~(2000) it was
concluded that the potential (gravitational) energy is larger than
the kinetic energy of the CMEs for relatively slow CMEs (which is
expected for failed eruptions), while the kinetic energy was found to
exceed the gravitational energy for a relatively fast CME (as
expected for eruptive CMEs). In the study of Aschwanden (2016)
the gravitational energy was found to make up a fraction of
$E_{grav}/E_{cme}=0.75\pm0.28$ of the total energy $E_{cme}
=E_{grav}+E_{kin}$, so that the most energetic CMEs 
(of GOES M and X-class flares) have a kinetic energy larger
than the gravitational energy, which was the case in 22\% of
the events. This low fraction is most likely
caused by the neglect of the solar wind drag force. 
Emslie et al.~(2012) estimated the CME kinetic energy in the
rest frame of the solar wind by subtracting 400 km s$^{-1}$
from the measured CME speed, which lowers the energy demand 
to overcome the flare-associated Lorentz force and thus increases the 
percentage of eruptive
CME events (compared with the percentage of failed eruptions).
In the study of Aschwanden (2017), the deceleration due to the
gravitational force was included in the dynamical model of initial CME 
acceleration, leading to a very small fraction of $\approx 2.3\%$ 
for failed eruptions. These are relatively low values compared
with the study of Cheng et al.~(2010), who found a fraction of
43\% for confined flare events. The lowest values of $\approx 2.3\%$
for failed CME eruptions may be a consequence of dynamic models
that over-estimate the CME velocity (Aschwanden 2017). In the 
present study we estimate a fraction of 40\%-83\% CME events
to be associated with ($>$M1.0 class) flares, depending on the 
chosen uncertainty of the time overlap 
($\Delta t \approx 0.7-4.0$ hrs; see Fig.~8), but is
largely consistent with earlier results, 
i.e., 43\% (Cheng et al.~2010) and 22\% (Aschwanden 2016). 

\subsection{	Aerodynamic Drag and Global Flare Energetics	}

How does the phenomenon of the aerodynamic drag force, which we
neglected in this series of statistical studies so far, effect
the global energy budget of a flare/CME event? 
In the study of Emslie et al.~(2012), the CME is estimated to
dissipate 19\% of the magnetic flare energy in the statistical
average. The total primary dissipated energy (by acceleration of 
nonthermal electrons and ions, as well as the
kinetic energy of CMEs) amount only to 25\% of the magnetic
energy in the study of Emslie et al~(2012), while the CME
kinetic energy (with the slow solar wind energy subtracted)
is estimated to consume 19\% of the available magnetic energy.
Since the effects of the slow solar wind has already been
corrected, no additional correction is needed to account
for the aerodynamic drag force, and thus the discrepancy 
in energy closure does not change, mostly caused by a
massive over-estimate of the magnetic energy, which was
estimated ad hoc to 30\% of the potential energy. 

In the study of Aschwanden (2017), energy closure is almost
reached (87\%$\pm$18\%), where the CMEs are estimated to dissipate
7\% of the available magnetic free energy. Including 
the energy supply by aerodynamic drag,  
the CME energy budget changes from 7\% to 4\% of the total flare
energy budget, and thus it just drops slightly in the 
energy closure from 87\% to 83\%. Hence there is no
dramatic change in the global flare energetics.

\subsection{	Coincidence of Flare and CME Starting Times	}

The association of flares and CMEs is fairly well established
by observing the initial rise of soft X-ray emission (such as
from a GOES light curve) and identifying a near-simultaneous 
EUV dimming, because these two time markers are produced
cospatially. It is more difficult to find the corresponding
flare-associated CME event from white-light observations (such
as with a height-time profile of the CME leading edge, because
the two associated phenomena are not cospatial. The time
delay between the GOES flare starting time and the first detection
with LASCO at $r \approx 3.0 R_{\odot}$ is 
$(t_1^{LASCO}-t_s^{GOES}) = 1.0 \pm 1.3$ hrs (Fig.~7a), 
during which multiple
flares can occur. One way to improve the simultaneity is to
extrapolate the LASCO height-time profile to the initial starting
height $r_s$, which indeed improves the coincidence to
$(t_s^{LASCO}-t_s^{GOES}) = 0.07 \pm 0.28$ hrs = $4 \pm 15$ minutes 
(Fig.~8a). The extrapolation from the first detection at $t_1^{LASCO}$
to the expected CME starting time $t_s^{LASCO}$, however, is 
model-dependent, and hence the timing uncertainties can amount
from $(t_s^{LASCO}-t_s^{GOES}) = 0.174$ hrs (Fig.~9a) to 
$(t_s^{LASCO}-t_s^{GOES}) = -1.659$ hrs (Fig.~9d).
More accurate starting time measurements could be achieved by
using occulter disks closer to the solar surface, such as
LASCO/C1, which unfortunately was disabled on June 1998. 

\subsection{	Estimating CME arrival times at Earth		}

An important parameter for space weather predictions is the estimated
propagation time from the solar CME site to the Earth at a distance
of 1 AU. In our study we compare the observed travel time
(Fig.~11, x-axis) with the predicted travel time (Fig.~11, y-axis)
based on the velocity profile $v(t)$ obtained from fitting the aerodynamic
drag model, which essentially is close to the travel time one obtains
from the slow solar wind speed of $w \approx 400$ km s$^{-1}$.
The comparison demonstrates that an accuracy of $\pm 23\%$ of the
observed travel time can be achieved, which translates for a range
of travel times ($\approx 35-87$) hrs to an uncertainty of
$\approx 8-20$ hrs. 

Our results compare favorably with other measurements.
Tucker-Hood et al.~(2014) report an average error of 22 hrs in the
predicted transit time, which exceeds the largest uncertainty of
our measurements. Kim et al.~(2007) compared 91 predictions of shocks
made with the empirical shock arrival model and found that 60\% of
the predicted travel times were within $\pm 12$ hrs.
McKenna-Lawlor et al.~(2006) found only 40\% of the cases
within $\pm 12$ hrs. One advantage of our method is that the
solar wind speed $w$ is measured from fitting the aerodynamic drag
model, so that no assumptions need to be made about the 
time-dependent variation of the slow solar wind. 

\section{	CONCLUSIONS 					}

Our motivation for this study is the role of the aerodynamic drag
force on the acceleration of CMEs, in the context of global energetics
of flares and CMEs. In previous studies on the energy closure and
partition in solar flares and CMEs we neglected this effect. Here
we investigate three data sets: one CME set that covers all (14,316) 
LASCO CME detections during the SDO era (2010-2017), one flare
data set with (576) GOES M- and X-class flares, and one set with (11)
interplanetary CMEs with known arrival times at Earth. We obtain
the following results:

\begin{enumerate}
\item{We apply two different forward-fitting models: (i) A
second-order polynomial fit based on the assumption of constant
acceleration during the propagation across the LASCO/C2 and C3  
coronagraph, and (ii) the aerodynamic drag model of Cargill (2004) and
Vrsnak et al.~(2013). Both are analytical models that can be fitted
to the observed height-time profiles $r(t)$ from LASCO and yield
either the acceleration constant $a$, or the ambient slow solar
wind speed $w$ and the drag coefficient $\gamma$. Both models fit the
data with an accuracy of $\approx 3\%$ in the ratio of modeled
to observed distances $r$. Both models can be applied to extrapolate
the starting time $t_s$ to the CME at a coronal base level 
$r_s = 1 R_{\odot}$ and to predict the arrival time of a CME at Earth.}

\item{The extrapolated starting times $t_s^{LASCO}$ are found to 
coincide with the flare starting time $t_s^{GOES}$ in soft X-rays
within $\pm 4$ hrs in 83\%, or within $\pm 0.7$ hrs in 40\%, which
implies that a fraction of 17\%-60\% of flare events have no
GOES $>$1 M class counterpart in LASCO-detected CMEs, possibly
representing failed eruptions or confined flare events. 
All LASCO-detected CMEs were
found to develop final speeds above the gravitational escape velocity,
latest after a distance of $r \gapprox 10 R_{\odot}$ or a travel time
of $t \gapprox 25$ hrs.} 

\item{The LASCO-detected CME events can be subdivided into two classes,
(i) one with dominant flare-associated acceleration in the lower corona
at heights of $r \lapprox 1.5 R_{\odot}$, inferred in 313 out of the
576 events, and (ii) one with dominant aerodynamic drag acceleration
in the upper corona of $r \approx (1.5-10.0) R_{\odot}$, identified
in 263 out of the 576 cases. The aerodynamic drag acceleration appears
to have an upper limit of CMEs kinetic energies at $E_{drag}\lapprox
2 \times 10^{32}$ erg, while the flare-associated acceleration can produce
CME kinetic energies up to $E_{flare} \approx 1.5 \times 10^{33}$ erg.
The ratio of the summed kinetic energies for the two acceleration processes
is $E_{flare}/E_{all} \approx 80\%$ for flare acceleration, and  
$E_{drag}/E_{all} \approx 20\%$ for the aerodynamic drag model,
so that $E_{flare}/E_{drag} \approx 4$.}

\item{The aerodynamic drag model predicts the velocity $v(t)$ of the
CME leading edges from the locations of LASCO detection all the way to
Earth, approaching asymptotically the solar wind speed at  
a distance of $r \gapprox 10 R_{\odot}$. For a subset of 11 events, 
for which the arrival times $\tau_{obs}$ at Earth are known, we predict
the arrival times $\tau^{pred}$ within an accuracy of $\approx 23\%$,
which translates into an uncertainty of 8-20 hrs.}

\item{For the global energetics of flare/CME events we found that CMEs
contribute in the average $\approx 7\%$ to the total energy budget,
for which we reached closure within $87\%\pm18\%$
(Aschwanden et al.~2017). Including the effects
of the aerodynamic drag, which boosts the CME kinetic energies in 
addition to the dissipated magnetic energies,
we find a correction of the estimated total energy by $\approx -4\%$,
which modifies energy closure from 87\% slightly downward to 83\%.}
\end{enumerate}

In summary, neglecting the aerodynamic drag does not modify the
overall energy budget by a large amount, i.e., the total dissipated
magnetic energy is reduced from a closure value of 87\% to 83\%, the
fraction of CME energies reduces from 7\% to $\approx 4\%$, but the 
kinetic energies in flare-accelerated CMEs are a factor of 
4 higher than the total kinetic energies transferred from the slow solar 
wind aerodynamic drag to the final CME kinetic energies. This preponderance
of flare-accelerated CME energies results from the inability of the 
aerodynamic drag to accelerate CMEs to larger kinetic energies than
$\lapprox 2 \times 10^{32}$ erg, while flares can produce CME kinetic
energies that are up to an order of magnitude higher.

\bigskip  
We acknowledge helpful discussions with Ian Richardson and Nariaki Nitta.
This CME catalog is generated and maintained at the CDAW Data Center 
by NASA and The Catholic University of America in cooperation with 
the Naval Research Laboratory. SOHO is a project of international 
cooperation between ESA and NASA. Support for the CDAW catalog
is provided by NASA/LWS and by the Air Force Office of
Scientific Research (AFOSR). 
This work was partially supported by NASA contracts NNX11A099G,
80NSSC18K0028, NNX16AF92G, and NNG04EA00C (SDO/AIA).

\clearpage

\section*{References} 
\def\ref#1{\par\noindent\hangindent1cm {#1}}

\ref{Aschwanden, M.J., Xu, Y., and Jing, J. 2014, ApJ 797, 50.
 	{\sl Global energetics of solar flares: I. Magnetic Energies}}
\ref{Aschwanden, M.J., Boerner, P., Ryan, D., Caspi, A., McTiernan, J.M., 
	and Warren, H.P., 2015, ApJ 802, 53.
 	{\sl Global energetics of solar flares: II. Thermal Energies}}
\ref{Aschwanden, M.J., O'Flannagain, A., Caspi, A., McTiernan, J.M., 
	Holman, G., Schwartz, R.A., and Kontar, E.P. 2016, ApJ 832, 27.
 	{\sl Global energetics of solar flares: III. Nonthermal Energies}}
\ref{Aschwanden, M.J. 2016, ApJ 831, 105.
	{\sl Global energetics of solar flares. IV. Coronal
	mass ejection energetics}}
\ref{Aschwanden, M.J., Caspi, A., Cohen, C.M.S., Holman, G.D., Jing, J., 
	Kretzschmar, M., Kontar, E.P., McTiernan, J.M., O'Flannagain, A., 
	Richardson, I.G., Ryan, D., Warren, H.P., and Xu,Y. 2017, ApJ 836, 17.
 	{\sl Global energetics of solar flares: V. Energy closure}}
\ref{Aschwanden, M.J. 2017, ApJ 847, 27.
	{\sl Global energetics of solar flares. VI. Refined
	energetics of coronal mass ejections.}}
\ref{Aschwanden, M.J. 2019, {\sl New Millennium Solar Physics},
	Astrophysics and Space Science Library Vol.~458, 
	ISBN 978-3-030-13954-4; New York: Springer.}
\ref{Bein, B.M., Berkebile-Stoiser, S., Veronig, A.M., et al.~(2011),
	ApJ 738, 191.
	{\sl Impulsive acceleration of coronal mass ejections.
	I. Statistics and coronal mass ejection source region characteristics}}
\ref{Cargill, P.J. 2004, SoPh 221, 135.
	{\sl On the aerodynamic drag force acting on interplanetary
	coronal mass ejections}} 
\ref{Chen,J. 1997, in {\sl Coronal Mass Ejections}, Geophys.Monogr.Ser. 99
	(eds. Crooker, N., Joslyn, J.A., and Feynman J., p.65, AGU, Washington
	DC. {\sl Coronal mass ejections: Causes and consequences - 
	A theoretical view}}
\ref{Cheng, X., Zhang, J., Ding, M.D., and Poomvises, W. 2010, ApJ 712, 752.
	{\sl A statistical study of the post-impulsive-phase acceleration of 
	flare-associated coronal mass ejections}}
\ref{Emslie, A.G., Kucharek, H., Dennis, B.R., Gopalswamy, N., Holman, G.D., 
	Share, G.H., Vourlidas, A., Forbes, T.G., Gallagher, P.T., Mason, G.M.,
	Metcalf, T.R., Mewaldt, R.A., Murphy, R.J., Schwartz, R.A., and 
	Zurbuchen, T.H. 2004, JGR (Space Physics), 109, A10, A10104.
 	{\sl Energy partition in two solar flare/CME events}}
\ref{Emslie, A.G., Dennis, B.R., Holman, G.D., and Hudson, H.S.,
 	2005, JGR (Space Physics), 110, 11103.
 	{\sl Refinements to Flare Energy Estimates - a Follow-up to "Energy 
	Partition in Two Solar Flare/CME Events"}}
\ref{Emslie, A.G., Dennis, B.R., Shih, A.Y., Chamberlin, P.C., Mewaldt, R.A., 
	Moore, C.S., Share, G.H., Vourlidas, A., and Welsch, B.T.
 	2012, ApJ 759, 71.
 	{\sl Global Energetics of Thirty-eight Large Solar Eruptive Events}}
\ref{Gopalswamy, N., Lara, A., Lepping, R.P. et al. 2000,
	GRL 27/2, 145.
	{\sl Interplanetary acceleration of coronal mass ejections}}
\ref{Gopalswamy, N., Yashiro, S., Kaiser, M.L., et al.~2001a,
	JGR 106, A12, 29219,
	{\sl Characteristics of coronal mass ejections associated with 
	long-wavelength type II radio bursts}}
\ref{Gopalswamy, N., Lara, A., Yashiro, S., et al.~2001b,
	JGR 106, A12, 29207.
	{\sl Predicting the 1-AU arrival times of coronal mass ejections}}
\ref{Gopalswamy, N., Yashiro, S., Liu, Y., et al. 2005a, JGR 110/A9,
	A09S15. {\sl Coronal mass ejections and other extreme 
	characteristics of the 2003 October-November solar eruptions}}
\ref{Gopalswamy, N., Lara, A., Manoharan, P.K. et al.~2005b,
	Adv.Space Res. 36/12, 2289.
	{\sl An empirical model to predict the 1-AU arrival of
	interplanetary shocks}}
\ref{Gopalswamy, N., Yashiro, S., Michalek, G. et al. 2009a,
	Earth, Moon, and Planets 104, 295.
	{\sl The SOHO/LASCO catalog}}
\ref{Gopalswamy, N., Akiyama, S., and Yashiro, S. 2009b,
	in Proc. {\sl Universal heliophysical processes},
	IAU Symp. 257, 283.
	{\sl Major solar flares without coronal mass ejections}}
\ref{Gopalswamy, N., Thompson, W.T., Davila, J.M., et al. 2009,
	SoPh 259, 227.
	{\sl Relation between type II bursts and CMEs inferred from
	STEREO observations}}
\ref{Gopalswamy, N., Yashiro, S., Michalek, G., Xie, H., M\"akel\"a, P.,
	Vourlidas, A., and Howard, R.A. 2010, Sun and Geosphere 5, 7.
	{\sl A catalog of halo coronal mass ejections from SOHO}}
\ref{{Gopalswamy, N., Nitta, N., Akiyama, S., et al. 2012,
	ApJ 744, 72.
	{\sl Coronal magnetic field measurement from EUV images made
	by the SDO}}}
\ref{Gopalswamy, N., M\"akel\"a, P., Xie, H., and Yashiro, S. 2013,
	Space Weather 11/11, 661.
	{\sl Testing the empirical shock arrival model using
	quadrature observations}}
\ref{Gopalswamy, N., Yashiro, S., Thakur, N., et al. 2016,
	ApJ 833, 216. {\sl The 2012 July 23 backside eruption:
	An extreme energetic particle event ?}}
\ref{Gopalswamy, N., M\"akel\"a, P., Yashiro, S., et al. 2017,
	J.Physics, Conf. Ser. 900, 012009.
	{\sl A hierarchical relationship between the fluence spectra
	and CME kinematics in large solar energetic particle events:
	A radio perspective}}
\ref{Hess,P. and Zhang, J. 2014, ApJ 792, 49. 
	{\sl Stereoscopic study of the kinematic evolution of a coronal mass
	ejection and its driven shock from the Sun to the Earth and the
	prediction of their arrival times}}
\ref{Iju, T., Tokumaru, M., and Fujiki, K. 2014, Solar Phys. 289, 2157.
	{\sl Kinematic Properties of Slow ICMEs and an Interpretation of 
	a Modified Drag Equation for Fast and Moderate ICMEs}}
\ref{Kay, C., dos Santos, L.F.G. and Opher, M. 2015, ApJ 801, L21.
	{\sl Constraining the Masses and the Non-radial Drag Coefficient 
	of a Solar Coronal Mass Ejection}}
\ref{Kilpua, E.K.J., Mierla, M., Rodriguez, L., Zhukov, A.N., Srivastava, N.,
	and West, M.J. 2012, Solar Phys. 279, 477.
	{\sl Estimating Travel Times of Coronal Mass Ejections to 1 AU Using 
	Multi-spacecraft Coronagraph Data}}
\ref{Kim, K.H., Moon, Y.J., and Cho, K.S. 2007, J. Geophys. Res. 112, A05104.
	{\sl Prediction of the 1-AU arrival times of CME-associated
	interplanetary shock propagation model}}
\ref{Lugaz, N. and Kintner, P. 2013, Solar Phys. 285, 281.
	{\sl Effect of Solar Wind Drag on the Determination of the Properties 
	of Coronal Mass Ejections from Heliospheric Images}}
\ref{Maloney, S.A. and Gallagher, P.T. 2010, ApJ 724, L127.
	{\sl Solar Wind Drag and the Kinematics of Interplanetary Coronal 
	Mass Ejections}}
\ref{Masson, S., Demoulin, P., Dasso, S., and Klein, K.L. 2012, A\&A 538, A32.
	{\sl The interplanetary magnetic structure that guides solar 
	relativistic particles}}
\ref{McKenna-Lawlor, S.M.P., Dryer, M., Kartalev, M.D., Smith, Z., Fry, C.D. et al.
	2006, J. Geophys. Res. 111, A11103.
	{\sl Near real-time predictions of the arrival at Earth of flare-related
	shocks during solar cycle 23}}
\ref{Michalek, G., Gopalswamy, N., Lara, A., and Manoharan, P.K. 2004, A\&A 423, 2.
	{\sl Arrival time of halo CMEs in the vicinity of the Earth}}
\ref{Michalek, G. 2012, Solar Phys. 276, 277.
	{\sl Dynamics of CMEs in the LASCO Field of View - Statistical 
	Analysis}}
\ref{Mittal, N., and Narain, U. 2015, Nat.Res.Inst.Aston.Geophys. 4, 100.
	{\sl On the arrival times of halo CME in the vicinity of the Earth}}
\ref{Press, W.H., Flannery, B.P., Teukolsy, S.A., and Vetterling, W.T.
	1986, Cambridge University Press, Cambridge.
	{\sl Numerical Recipes. The Art of Scientific Computing}}
\ref{Sachdeva, N., Subramanian, P., Colaninno, R., and Vourlidas, A.
	2015, ApJ 809, 158.
	{\sl CME Propagation: Where does Aerodynamic Drag 'Take Over'?}}
\ref{Sachdeva, N., Subramanian, P., Vourlidas, A., and Bothmer, C.
	2017, Solar Phys. 292, 118.
	{\sl CME Dynamics Using STEREO and LASCO Observations: 
	The Relative Importance of Lorentz Forces and Solar Wind Drag}}
\ref{Shen, F., Wu, S.T., Feng, Z., and Wu, C.C.
	2012, JGR 117, A11, CiteID A11101.
	{\sl Acceleration and deceleration of coronal mass ejections 
	during propagation and interaction}}
\ref{Song, W.B. 2010, Solar Phys. 261, 311.
	{\sl An analytical model to predict the arrival time of
	interplanetary CMEs}}
\ref{Subranmanian, P., Lara, A., and Borgazzi, A. 2012,
	GRL 39/9, CiteID L19107.
	{\sl Can solar wind viscous drag account for coronal mass ejection 
	deceleration?}}
\ref{Temmer, M. and Nitta, N.V. 2015, Solar Phys. 290, 919.
	{\sl Interplanetary propagation behavior of the fast coronal
	mass ejection on 23 July 2012}}
\ref{Tucker-Hood, K., Scott, C., Owens, M., Jackson, D., Barnard, L.,
	Davies, J.A., Crothers, S., Lintott, C., et al.
	2014, Space Weather, 10.1002/2014SW001106.
	{\sl Validation of a priori CME arrival predictions made using
	real-time heliospheric imager observations}}
\ref{Verbeke, C., Mays, M.L., Temmer, M., Bingham, S., Steenburgh, R.,
	Umbovic, M.N., Nez, M.N., Jian K,J., Hess, P., Wiegard, C., 
	Tatakishvili, A., and Andries, J. 2019, eprint-archive/
	{\sl Benchmarking CME arrival time and impact: Progress on 
	metadata, metric, and events}} 
\ref{Vourlidas, A., Subramanian, P., Dere, K.P., and Howrd, R.A.
	2000, ApJ 534, 456.
	{\sl Large-angle spectrometric coronagraph measurements of the
	energetics of coronal mass ejections}}
\ref{Vrsnak, B., and Gopalswamy, N. 2002, JGR (Space Physics) 107, A2, CiteID 1019.
	{\sl Influence of the aerodynamic drag on the motion of
	interplanetary ejecta}}
\ref{Vrsnak, B., Vrbanec, D., and Calogovic, J. 2008, A\&A 490, 811.
	{\sl Dynamics of coronal mass ejections. The mass-scaling of the
	aerodynamic drag}}
\ref{Vrsnak, B., Zic, T., Falkenberg, T.V., M\"ostl, C., Vennerstrom, S.,
	and Vrbanec, D. 2010, A\&A 512, A43.
	{\sl The role of aerodynamic drag in propagation of interplanetary 
	coronal mass ejections}}  
\ref{Vrsnak, B., Zic, T., Vrbanec, D., et al. 2013, SoPh 285, 295.
	{\sl Propagation of interplanetary coronal mass ejections:
	The Drag-based model.}}
\ref{Vrsnak, B., Temmer, M., Zic, T., Tatakishvili, A., Dumbovic M.,
	M\"ostl, C., Veronig, A.M., Mays, M.L., and Odstrcil, D.
	2014, ApJSS 213, 21.
	{\sl Heliospheric Propagation of Coronal Mass Ejections: 
	Comparison of Numerical WSA-ENLIL+Cone Model and Analytical 
	Drag-based Model}}
\ref{Yashiro, S., Gopalswamy, N., Akiyama, S., et al.~2005,
	JRG 110, A12, A12S05.
	{\sl Visibility of coronal mass ejections as a function of flare
	location and intensity}}
\ref{Yashiro, S., Michalek, G., and Gopalswamy, N. 2008,
	Ann.Geophys. 26, 3103.
	{\sl A comparison of coronal mass ejections identified
	by manual and automatic methods}}
\ref{Zic, T., Vrsnak, B., and Temmer, M. 2915, ApJSS 218, 32.
	{\sl Heliospheric Propagation of Coronal Mass Ejections: 
	Drag-based Model Fitting}}

\clearpage


\begin{table}
\tabletypesize{\normalsize}
\caption{Statistics of CME parameters (mean, standard deviation, median)
for 14,316 eruptive CME events detected with LASCO/SOHO during 2010-2017.}
\medskip
\begin{tabular}{lll}
\hline
Parameter &
Mean and standard dev. & 
Median \\
\hline
Starting height $r_s$			& $1.7\pm1.4$ $R_{\odot}$ & 1.2 $R_{\odot}$ \\
Height of first LASCO detection $r_1$   & $3.0\pm0.8$ $R_{\odot}$ & 2.7 $R_{\odot}$ \\
Height of last LASCO detection $r_2$    & $10.3\pm6.4$ $R_{\odot}$   & 8.1 $R_{\odot}$ \\
Starting velocity $v_s$			& $482\pm1294$ km/s & 202 km/s \\
Velocity at first LASCO detection $v_1$ & $320\pm283$ km/s & 284 km/s \\
Velocity at last LASCO detection $v_2$  & $368\pm198$ km/s & 326 km/s \\
Slow solar wind speed $w$               & $472\pm414$ km/s & 405 km/s \\
Acceleration $|a|$			& $0.013\pm0.029$ km/s$^{-2}$ & $0.005$ km/s$^{-2}$ \\
LASCO detection delay $t_1-t_s$ 	& $1.0\pm1.3$ hrs & 0.9 hrs \\
LASCO detection duration $t_2-t_1$ 	& $4.3\pm3.7$ hrs & 3.2 hrs \\
Aerodynamic drag coefficient $\gamma$   & $(2.6\pm3.3) \times 10^{-7}$ cm$^{-1}$ & $1.3 \times 10^{-7}$ cm$^{-1}$  \\
Accuracy of constant-acceleration model $\sigma_{CA}$ & $2.7\%\pm2.7\%$ & 2.4\% \\
Accuracy of aerodynamic drag model      $\sigma_{AD}$ & $2.9\%\pm2.5\%$ & 2.5\% \\
Number of observed LASCO images	$n_t$	& $23\pm17$ & 19 \\	
Average cadence 			& $0.21 \pm 0.07$ hrs & 0.20 hrs = 12 min \\	
Starting time delay $(t_s^{LASCO}-t_s^{GOES})$ & 0.07$\pm$0.27 hrs & -0.06 hrs \\
\hline
\end{tabular}
\end{table}

\begin{table}
\tabletypesize{\normalsize}
\caption{Observed and predicted arrival time at Earth for 14 eruptive CME events,
(data extracted from Interplanetary Coronal Mass Ejection web site,
provided by Ian Richardson and Hillary Cane), and based on an empirical
correction factor of $q_{corr}=0.81$ due to velocity projection effects.}
\medskip
\begin{tabular}{rrrrrrrr}
\hline
\# & Start time            & Arrival time        &Velocity& Solar wind& Observed & Predicted & Ratio\\
   & GOES                  & ICME at Earth       & $v_2$  & speed $w$ & delay    & delay     &      \\
   & (UT)                  & (UT)                & (km/s) & (km/s)    & (hrs)    & (hrs)     &      \\
\hline
  12 & 2011-02-15T01:44:00 & 2011-02-18T01:30:00 &    581 &    436 &     71 &     69 &    0.961 \\
  54 & 2011-08-02T05:19:00 & 2011-08-04T21:53:00 &    611 &    438 &     64 &     67 &    1.038 \\
  58 & 2011-08-04T03:41:00 & 2011-08-05T17:51:00 &   1110 &    467 &     38 &     47 &    1.231 \\
  66 & 2011-09-06T22:12:00 & 2011-09-09T12:42:00 &    565 &    425 &     62 &     64 &    1.024 \\
  98 & 2011-10-02T00:37:00 & 2011-10-05T07:36:00 &    264 &    389 &     78 &    103 &    1.304 \\
 115 & 2011-11-09T13:04:00 & 2011-11-12T05:59:00 &    789 &    448 &     64 &     55 &    0.847 \\
 147 & 2012-03-07T00:02:00 & 2012-03-08T11:03:00 &   2405 &    487 &     35 &     22 &    0.628 \\
 273 & 2013-04-11T06:55:00 & 2013-04-13T22:54:00 &    775 &    439 &     63 &     57 &    0.891 \\
 409 & 2014-02-04T01:16:00 & 2014-02-07T17:05:00 &    488 &    419 &     87 &     74 &    0.843 \\
 421 & 2014-02-12T06:54:00 & 2014-02-15T13:16:00 &    432 &    818 &     78 &     64 &    0.817 \\
 504 & 2014-09-10T17:21:00 & 2014-09-12T15:53:00 &    955 &    403 &     46 &     64 &    1.375 \\
mean &                     &                     &        &        &        &        & 1.00$\pm$0.23 \\
\hline
\end{tabular}
\end{table}


\begin{figure}
\centerline{\includegraphics[width=1.0\textwidth]{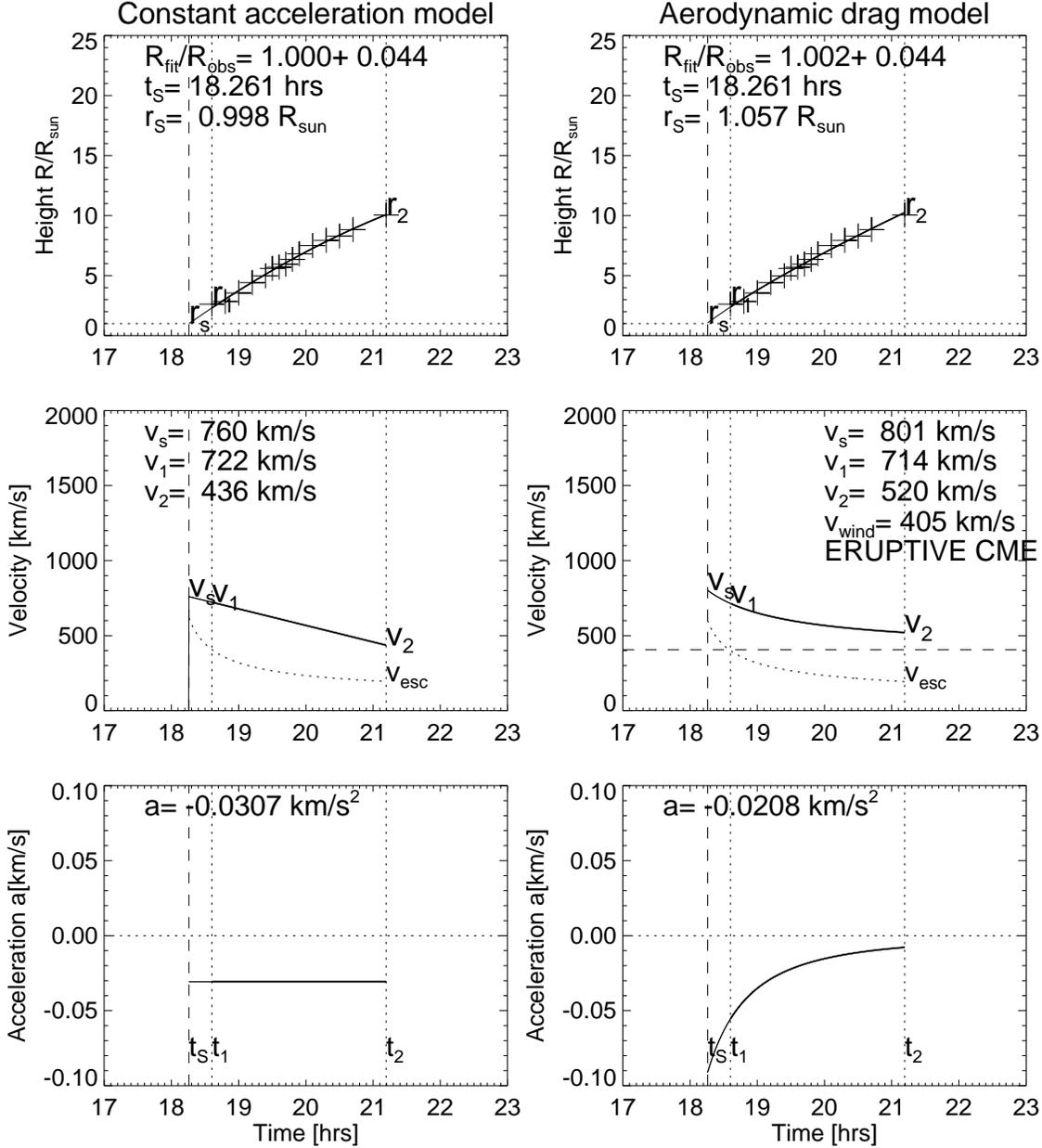}}
\caption{The height-time profile $r(t)$ (top panels), the velocity 
profile $v(t)$ (middle panels), and the acceleration profile $a(t)$ 
(bottom panels) of two CME kinematic models, the constant-acceleration
model (left panels), and the aerodynamic drag model (right panels),
showing an example of CME decelaration ($v_2 < v_1$).
The observed data points (crosses in top panels) are detected during
the time interval $[t_1, t_2]$, while the approximate starting time
$t_s$ is constrained by the initial height $r_s=r(t=t_s)$.}
\end{figure}

\begin{figure}
\centerline{\includegraphics[width=1.0\textwidth]{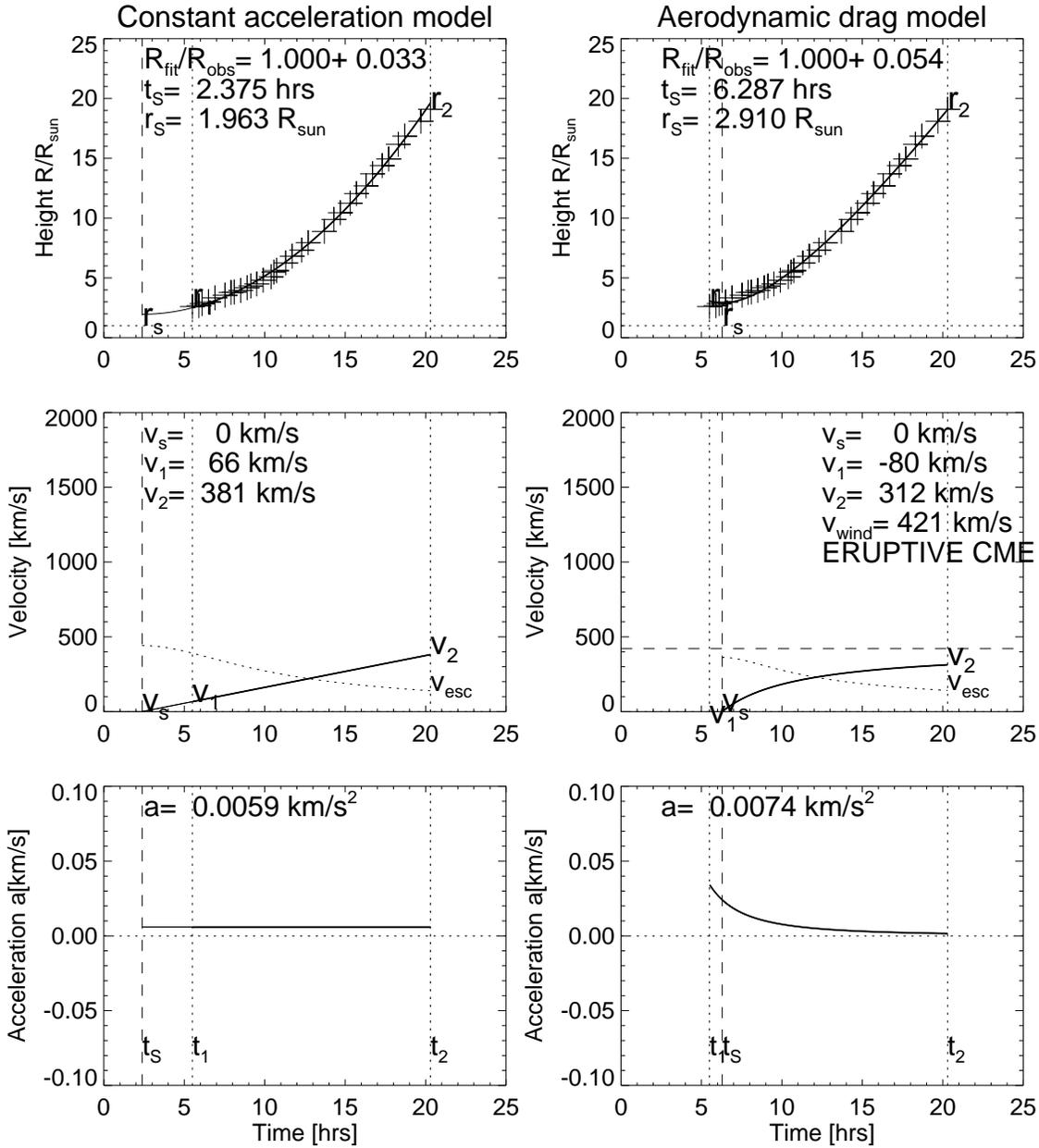}}
\caption{An example of a CME with acceleration is shown, $v_2 > v_1$.
Representation otherwise similar to Fig.~1.}
\end{figure}

\begin{figure}
\centerline{\includegraphics[width=1.0\textwidth]{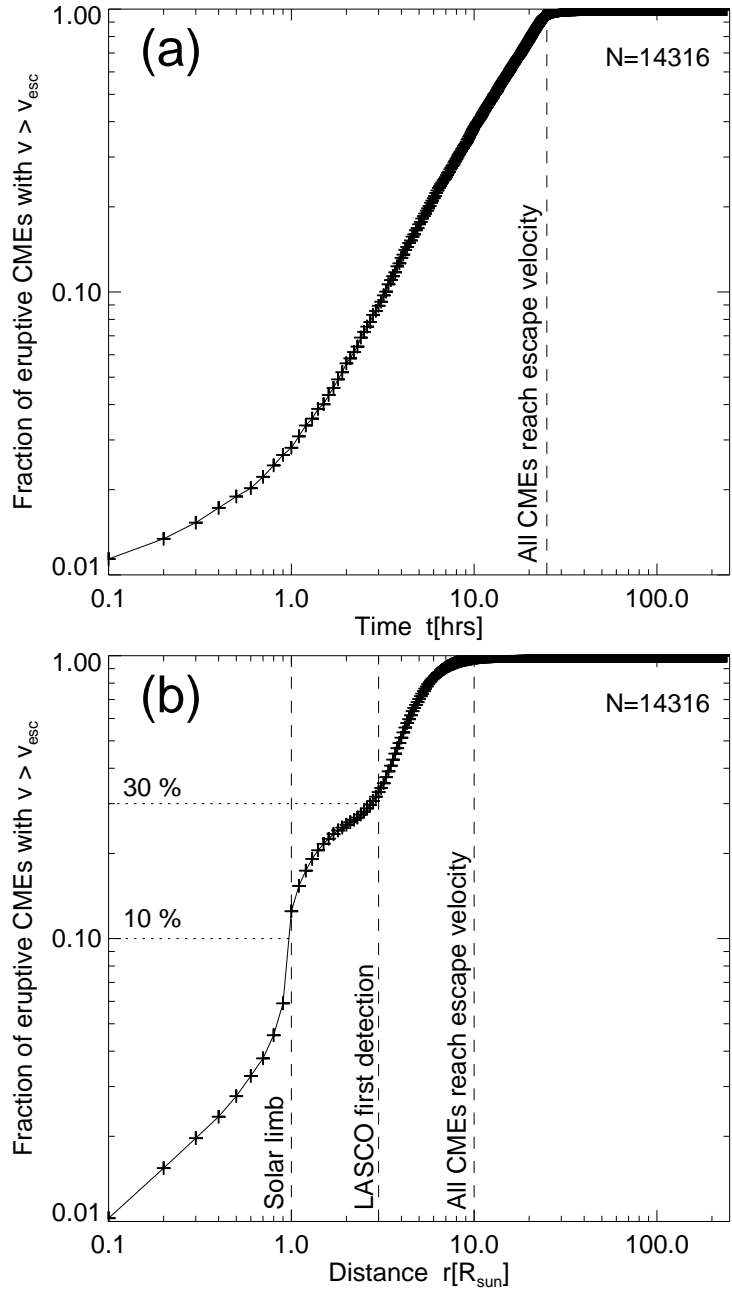}}
\caption{Fraction $q_t=N_{esc}(t)/N_{all}$ of eruptive 
CME events that exceed the escape velocity, as a function of 
the travel time, with $v(t) > v_{esc}(t)$ (a), and as a
function of the travel distance, with $v(r) > v_{esc}(r)$ (b).}
\end{figure}

\begin{figure}
\centerline{\includegraphics[width=1.0\textwidth]{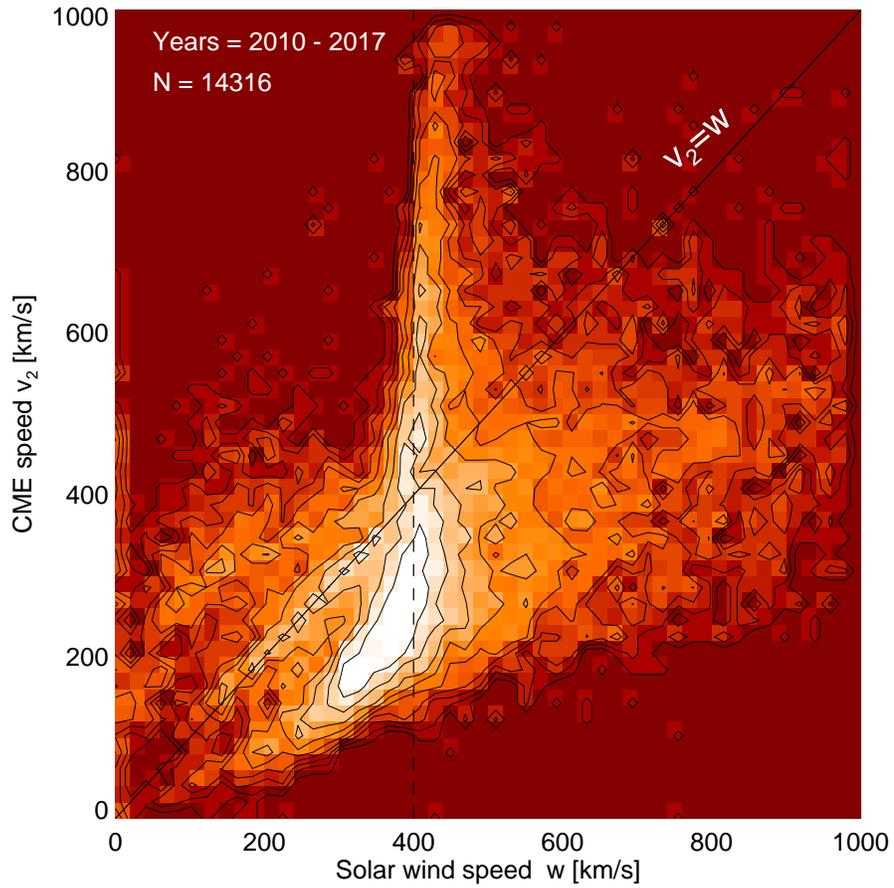}}
\caption{Comparison of final CME speed $v_2$ (at the last
detection with LASCO) with the solar wind speed $w$, for
all CME events.  Equivalence of CME speed $v_2$ and wind speed $w$
is indicated with a diagonal line, while the vertical dashed
line indicates a slow solar wind speed of $w=400$ km.}
\end{figure}

\begin{figure}
\centerline{\includegraphics[width=1.0\textwidth]{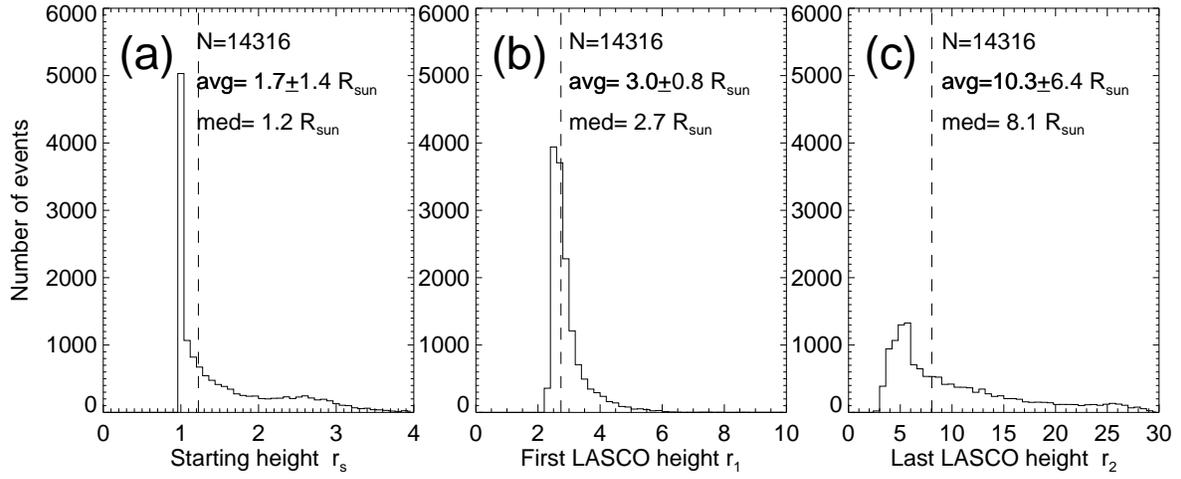}}
\caption{Distributions of CME starting heights $r_s$ (a),
heights $r_1$ of first LASCO detection (b),
and heights $r_2$ of last LASCO detection (c).
The median values of the distributions are marked with 
a vertical dashed line.}
\end{figure}

\begin{figure}
\centerline{\includegraphics[width=1.0\textwidth]{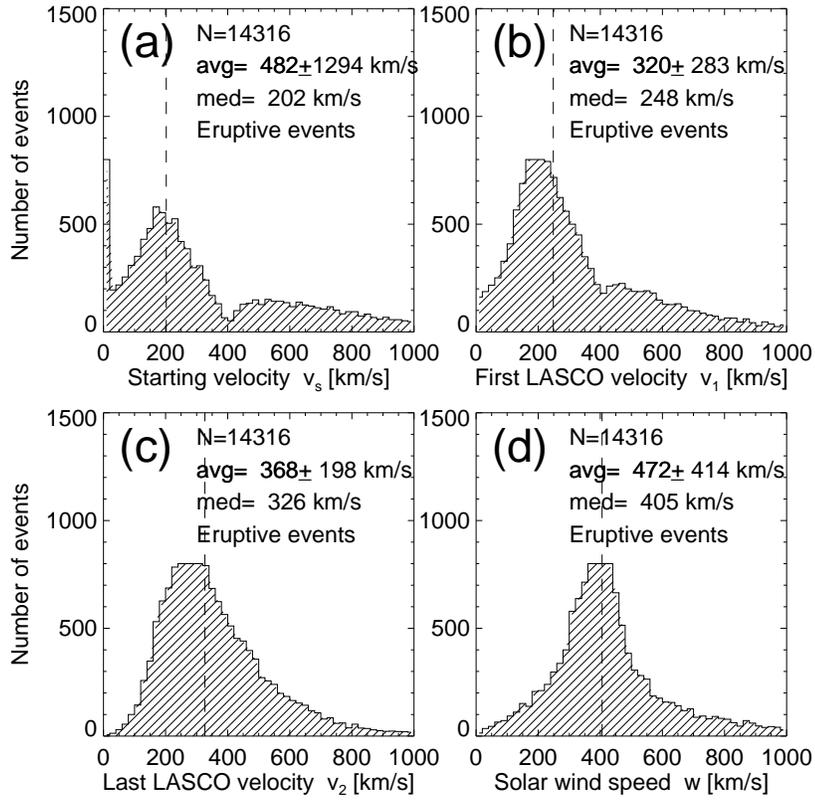}}
\caption{Distributions of CME starting velocities $v_s$ (a),
velocities $v_1$ of LASCO first detection (b),
velocities $v_2$ of LASCO last detection (c), and
ambient solar wind speed $w$ (d).
The median values of the distributions are marked with a 
vertical dashed line.}
\end{figure}

\begin{figure}
\centerline{\includegraphics[width=1.0\textwidth]{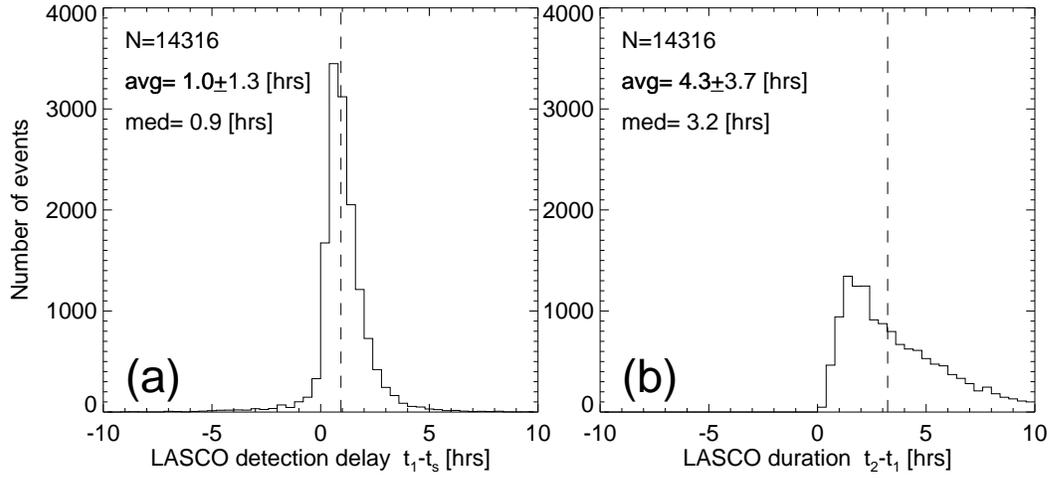}}
\caption{Distributions of LASCO detection delays (a), 
and durations (b) of CME detection detected in LASCO C2,C3 
field-of-views. The medians of the distributions are 
indicated with vertical dashed lines.}
\end{figure}

\begin{figure}
\centerline{\includegraphics[width=1.0\textwidth]{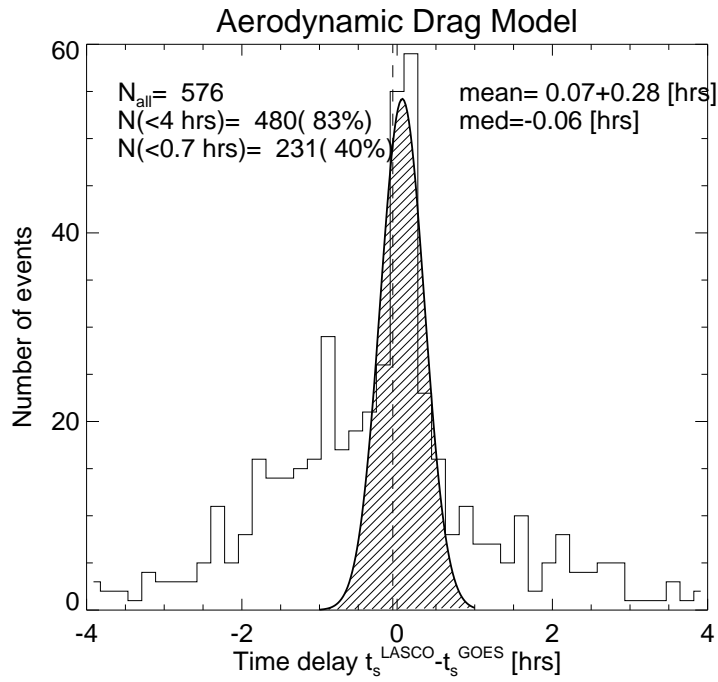}}
\caption{Distribution of time delays between the LASCO
extrapolated starting time and the GOES flare start time,
$\tau=t_s^{LASCO}-t_s^{GOES}$, with a Gaussian fit in the
core of the distribution. The median is indicated with a
vertical dashed line.}
\end{figure}

\begin{figure}
\centerline{\includegraphics[width=0.9\textwidth]{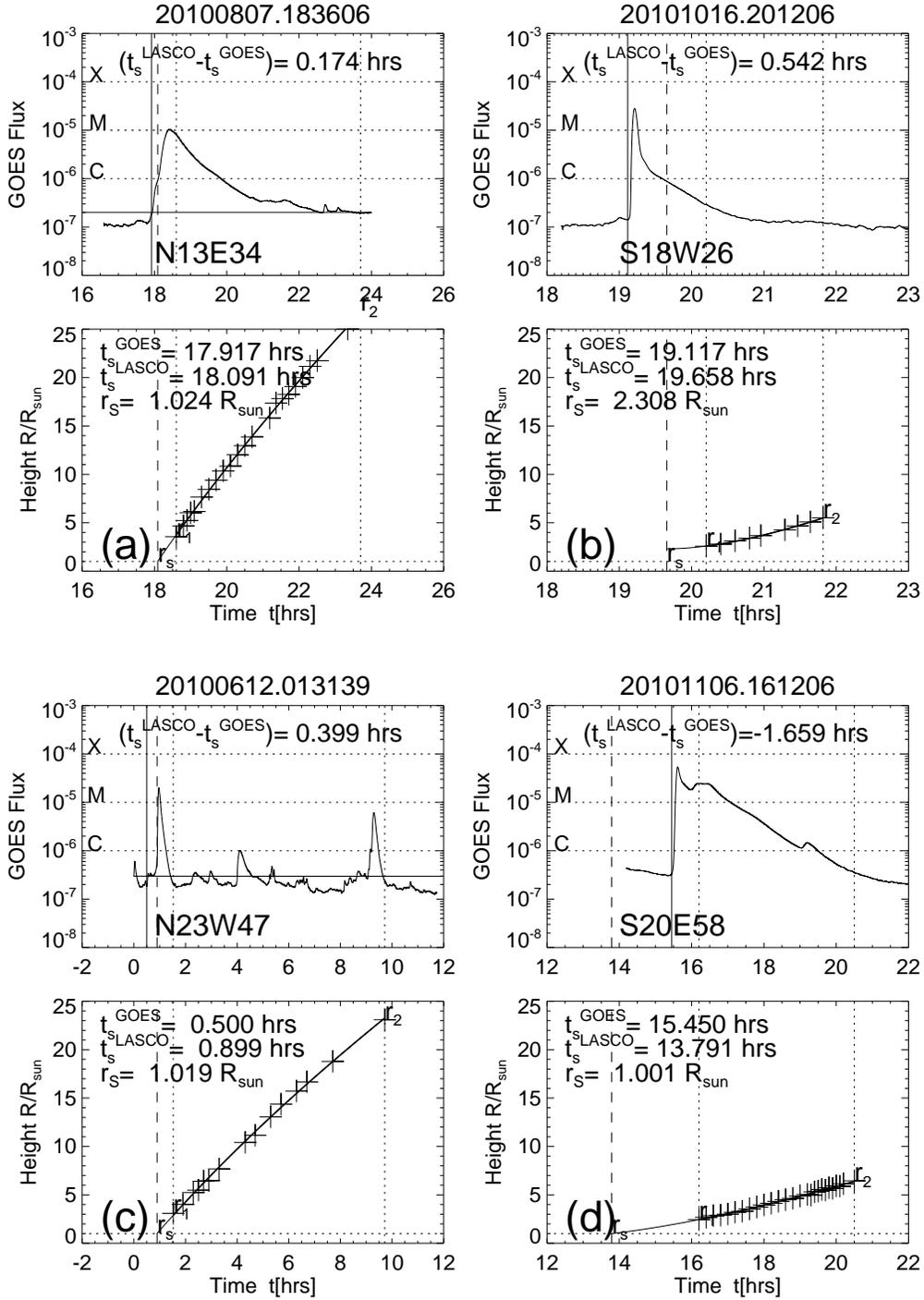}}
\caption{Four examples of flares with GOES flux time profiles
and height-time profiles $R/R_{\odot}$ are shown. The fitted
range is demarcated with vertical dotted lines and cross
symbols, the GOES starting time $t_s^{GOES}$ with a vertical solid line,
and the extrapolated CME starting time $t_s^{LASCO}$ at
a height of $r_s \approx 1 R_{\odot}$ with
a vertical dashed line. The GOES and LASCO starting times
coincide within the indicated fraction of hours, 
$(t_s^{LASCO}-t_s^{GOES})$. The heliographic flare location
is indicated in the bottom left of the GOES panels.}
\end{figure}

\begin{figure}
\centerline{\includegraphics[width=1.0\textwidth]{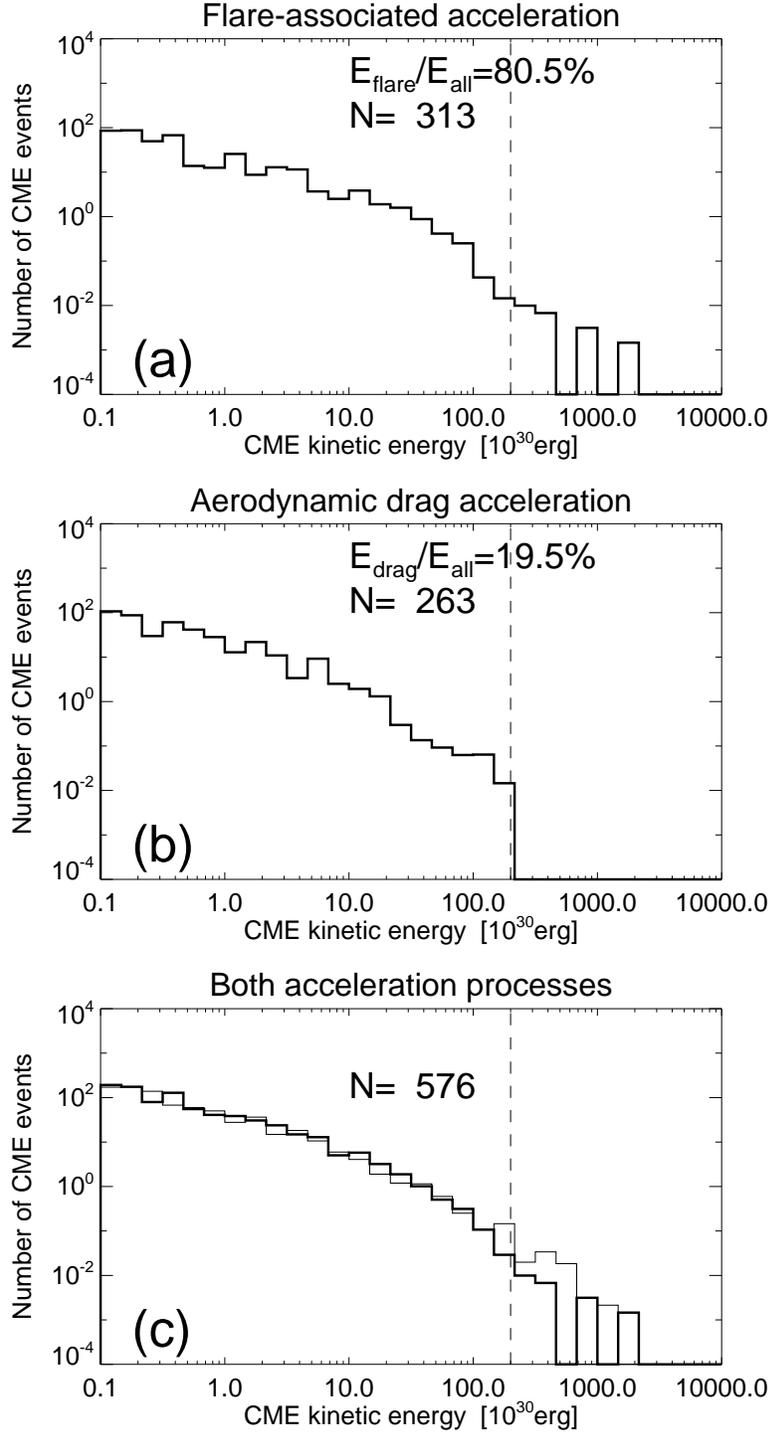}}
\caption{The logarithmic distribution of CME kinetic energies 
(histograms with thick linestyle)
for flare-associated acceleration events (a), aerodynamic
drag acceleration events (b), and the sum of both event types.
The fraction of the total CME energies integrated over the entire
distributions are indicated, along with the number of events.
For comparison, the distribution of a previous study 
(Aschwanden 2017) is shown also (histogram with thin linestyle
in bottom panel (c). Note that aerodynamic drag acceleration
shows an upper limit of $E\lapprox 2 \times 10^{32}$ erg (vertical
dashed line).}
\end{figure}

\begin{figure}
\centerline{\includegraphics[width=1.0\textwidth]{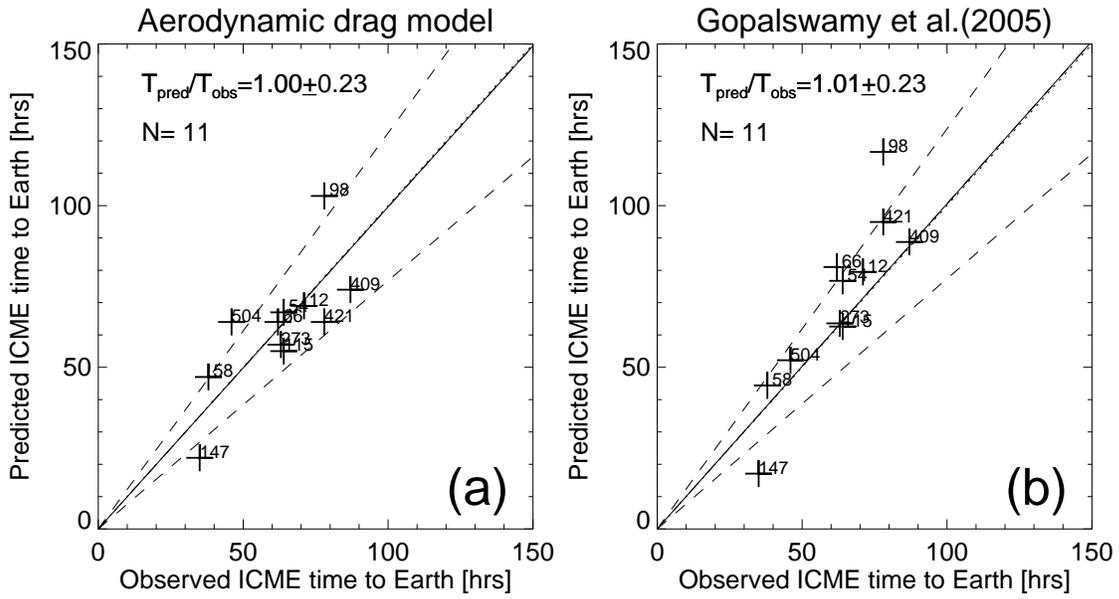}}
\caption{(a) The predicted ICME travel time from the Sun to Earth
as a function of the observed travel time for 11 ICME events,
normalized the empirical factor $q_{corr}=0.81$. The resulting
average ratio is $T_{pred}/T_{obs}=1.00\pm0.23$, which
implies that the ICME travel time can be predicted with an
accuracy of $\approx 23\%$. (b) Using prediction from empirical
formula of Gopalswamy et al.~(2005). Note the identical values
for the standard deviation.}
\end{figure}

\end{document}